\documentclass{jfm}
\usepackage{relsize}
\usepackage{caption}
\usepackage{subcaption}
\usepackage{graphicx}
\usepackage{natbib}
\usepackage{mathrsfs}
\usepackage{setspace}
\usepackage{amsmath,mathtools}
\usepackage[usenames,dvipsnames]{color} % comments and color
\usepackage{ifthen}

\usepackage{tabularx}
\usepackage{longtable}
\usepackage{listings}
\usepackage{booktabs}
\usepackage{float}
\usepackage{makecell}
\usepackage{epstopdf}
\usepackage{pifont} %%This package is for \ding
\usepackage{wasysym}%%This package is for \Circle 
\usepackage{tikz}
\usepackage{upgreek}
\usepackage[normalem]{ulem}

\usepackage{flafter}  % put this in your pre-amble

\newboolean{showcomments}
\setboolean{showcomments}{true}

\newcommand{\dennice}[1]{  \ifthenelse{\boolean{showcomments}}
{\textcolor{red}{(Dennice says:  #1)}}{}}

\def\mline{\vrule width4pt height2.5pt depth -2pt}
\def\dashed{\mline\hskip3.5pt\mline\thinspace}
\def\dashdot{\mline\ \bdot\,\  \mline\thinspace}

\def\bdot{\raise.2em\hbox to .15em{.}}

\newtheorem{thm}{Theorem}
\newtheorem{definition}[thm]{Definition} % reset theorem numbering for each chapter

\newtheorem{corollary}[thm]{Corollary}
\newenvironment{myproof}{\paragraph{\textbf{Proof:}}}{\hfill$\square$}

\usepackage{lmodern,pdftexcmds,amsmath}

\DeclareMathAlphabet{\mathsfbi}{OT1}{\sfdefault}{bx}{sl}
\DeclareMathVersion{sfletters}
\SetSymbolFont{letters}{sfletters}{OML}{ntxsfmi}{b}{it}

\makeatletter
\newcommand{\mathbfsbilow}[1]{%
  \text{\mathversion{sfletters}$\m@th#1$}%
}
\makeatother

\usepackage{etoolbox}

\graphicspath{{figures/}} % change it accordingly!

\begin{document}

\shorttitle{Structured input-output analysis of stratified plane Couette flow} %for header on odd pages
\shortauthor{C. Liu, C. P. Caulfield, D. F. Gayme} %for header on even pages

\title{Structured input-output analysis of stably stratified plane Couette flow}

\author
 {
 Chang Liu\aff{1}
  \corresp{\email{cliu124@alumni.jh.edu}},
  Colm-cille P. Caulfield \aff{2,3}
  Dennice F. Gayme \aff{1}
  %H. - C. Smith\aff{1, 2}
  %\and 
  %J. Q.  Public\aff{2}
  }

\affiliation
{
\aff{1}
Department of Mechanical Engineering, Johns Hopkins University, Baltimore, MD 21218, USA
\aff{2} BP Institute, University of Cambridge, Cambridge CB3 0EZ, United Kingdom
\aff{3} Department of Applied Mathematics and Theoretical Physics, University of Cambridge,
Cambridge CB3 0WA, United Kingdom
}

\maketitle

\begin{abstract}
We employ a recently introduced structured input-output analysis (SIOA) approach to analyze streamwise and spanwise wavelengths of flow structures in stably stratified plane Couette flow. In the low-Reynolds number ($Re$) low-bulk Richardson number ($Ri_b$) spatially intermittent regime, we demonstrate that SIOA predicts high amplification associated with wavelengths corresponding to the characteristic oblique turbulent bands in this regime. SIOA also identifies quasi-horizontal flow structures resembling the turbulent-laminar layers commonly observed in the high-$Re$ high-$Ri_b$ intermittent regime. An SIOA across a range of $Ri_b$ and $Re$ values suggests that the classical Miles-Howard stability criterion ($Ri_b\leq 1/4$) is associated with a change in the most amplified flow structures when Prandtl number is close to one ($Pr\approx 1$). However, for $Pr\ll 1$, the most amplified flow structures are determined by the product $PrRi_b$. For $Pr\gg 1$, SIOA identifies another quasi-horizontal flow structure that we show is principally associated with density perturbations. We further demonstrate the dominance of this density-associated flow structure in the high $Pr$ limit by constructing analytical scaling arguments for the amplification in terms of $Re$ and $Pr$ under the assumptions of unstratified flow (with $Ri_b=0$)  and streamwise invariance. 

%%------------Abstract for JFM html
%We employ a recently introduced structured input-output analysis (SIOA) approach to analyze streamwise and spanwise wavelengths of flow structures in stably stratified plane Couette flow. In the low-Reynolds number (Re) low-bulk Richardson number (Ri<sub>b</sub>) spatially intermittent regime, we demonstrate that SIOA predicts high amplification associated with wavelengths corresponding to the characteristic oblique turbulent bands in this regime. SIOA also identifies quasi-horizontal flow structures resembling the turbulent-laminar layers commonly observed in the high-Re high-Ri<sub>b</sub> intermittent regime. An SIOA across a range of Ri<sub>b</sub> and Re values suggests that the classical Miles-Howard stability criterion (Ri<sub>b</sub>≤1/4) is associated with a change in the most amplified flow structures when Prandtl number is close to one (Pr≈1). However, for Pr«1, the most amplified flow structures are determined by the product PrRi<sub>b</sub>. For Pr»1, SIOA identifies another quasi-horizontal flow structure that we show is principally associated with density perturbations. We further demonstrate the dominance of this density-associated flow structure in the high Pr limit by constructing analytical scaling arguments for the amplification in terms of Re and Pr under the assumptions of unstratified flow (with Ri<sub>b</sub>=0)  and streamwise invariance. 

\end{abstract}

\section{Introduction}
\label{sec:stratified_introduction}

Statically stable density stratification in wall-bounded shear flows plays an important role in many industrial and environmental applications, e.g., in cooling equipment \citep{zonta2018stably}, and the turbulent boundary layers governing atmospheric and oceanic flows \citep{vallis2017atmospheric,pedlosky2013geophysical}. In the atmospheric boundary layer, stable stratification arising from strong ground cooling effects is of particular importance at night~\citep{nieuwstadt1984turbulent,mahrt1999stratified,mahrt2014stably} and near the polar region \citep{grachev2005stable}. At the ocean floor, stable density stratification is known to influence the boundary layer thickness \citep{weatherly1978structure,lien2004turbulence}.  

(Stably) stratified plane Couette flow (PCF) is a canonical model for stratified wall-bounded shear flow. When the density as well as the velocity is maintained at different values
at the two horizontal boundary planes, with gravity acting vertically, stratified PCF has the added benefit (as defined more precisely below) that a natural bulk Richardson number $Ri_b$ can be defined, capturing the relative
significance of the imposed stratification and shear. Furthermore, unstratified PCF
has no linear instability for any Reynolds number ($Re$, again defined more precisely below) \citep{romanov1973stability}, and yet is observed to  transition  at Reynolds numbers as low as $Re=360\pm 10$ \citep{tillmark1992experiments}. Stratified PCF is a convenient model flow
for investigating the effect of stable stratification on transition dynamics \citep{deusebio2015intermittency}.

Stable stratification provides a restoring buoyancy force inhibiting vertical motion \citep{turner1979buoyancy,davidson2013turbulence}. Thus, transition to turbulence in stably  stratified PCF typically occurs at a higher Reynolds number than unstratified PCF; see e.g., \citep{deusebio2015intermittency,eaves2015disruption,olvera2017exact,deguchi2017scaling}. In the transitional regime, both stratified PCF and unstratified PCF exhibit spatial intermittency; i.e., the coexistence of laminar and turbulent regions. In the relatively low-$Re$ low-$Ri_b$ intermittent regime, the spatial intermittency in stratified PCF is characterized by oblique turbulent bands \citep{deusebio2015intermittency,taylor2016new} at least qualitatively similar to the unstratified PCF \citep{prigent2003long,duguet2010formation} with a very large channel size ($\sim O(100)$ times of channel half-height). In the high-$Re$ high-$Ri_b$ intermittent regime, flow structures are instead characterized by turbulent and laminar layers over the vertical direction due to the strong effect of buoyancy \citep{deusebio2015intermittency}. This spatial intermittency directly imposes challenges for the computation of averaged measurements of flow behavior (such as the efficiency of mixing or the dissipation rate), and thus understanding underlying mechanisms is important for the parameterization of turbulence properties, in particular the irreversible mixing in stratified flows \citep{caulfield2020open,caulfield2020layering}.

The existence of a unique critical Richardson number  that separates flow into laminar and turbulent regimes is  questionable, to put it mildly \citep{galperin2007critical,andreas2002parameterizing}. A threshold value close to $1/4$ is supported by some field measurements \citep{kundu1991evidence} and experiments \citep{rohr1988growth}, although other field measurements reported sustained turbulence in flows with Richardson numbers $\simeq 1$ \citep{lyons1964critical}. More recently, increasing evidence has been found that vertically sheared stably stratified flow appears to self-organize to maintain an appropriately defined Richardson number near $1/4$, both in field observations \citep{smyth2013marginal,smyth2019self} and in simulations \citep{salehipour2018self}. This threshold value of $1/4$ also appears in the classical `Miles-Howard' theorem \citep{miles1961stability,howard1961note}, which provides a necessary condition for linear instability in inviscid, non-diffusive steady parallel flow. In particular, it states that instability requires the local or gradient Richardson number must be less than $1/4$ somewhere. Therefore, it is of interest to consider stratified PCF as a well-controlled sheared stratified flow to investigate whether some kind of self-organized criticality and/or marginal stability naturally emerges in a viscous and diffusive flow.

The Prandtl number ($Pr=\nu/\kappa$, where $\nu$ is the kinematic viscosity and $\kappa$ is the diffusivity of the density field) plays a perhaps unsurprisingly important role in determining flow structures. For example, for sufficiently small $Pr$, flows with the same value of the product $Pr Ri_b$ develop the same  averaged vertical density profile \citep{langham2020stably}. This observation that the product $PrRi_b$  determines flow behavior in the low Prandtl number limit is widely observed in stratified shear flows; see e.g., \citep{lignieres1999small,garaud2015stability,garaud2017turbulent,garaud2021journey}. 
Conversely, in the high Prandtl number regime, exact coherent structures in stratified PCF \citep{langham2020stably} show that a nearly uniform density region forms near the channel center, and the influence of bulk Richardson number on the averaged properties of the flow are significantly reduced. Moreover, \cite{taylor2017multi} proposed a multi-parameter criterion for the formation of a `staircase' in the density distribution (i.e. a distribution with relatively deep  `layers' of nearly uniform density separated by relatively thin interfaces of enhanced density gradient)  which suggests that staircase formation is actually favored for a large Prandtl number \citep{taylor2017multi}. The sharpness of the density interfaces also appears to increase as the Prandtl number increases \citep{zhou2017diapycnal}. In addition,  increasing the Prandtl number has a larger influence on the mean density profiles than on the mean velocity profiles \citep{zhou2017self}. 

The oblique turbulent bands observed in the intermittent regime of stratified PCF \citep{deusebio2015intermittency,taylor2016new} require a very large channel size to  accommodate them fully, which poses challenges for both simulations and experiments. The three different flow parameters of interest, $Re$, $Ri_b$ and $Pr$ also lead to computational challenges in exploring the full range of flow regimes. To overcome these challenges to direct numerical simulation (DNS), we use an  input--output (resolvent) analysis based approach. Such methods, built upon the spatio-temporal frequency response, have been widely employed in unstratified wall-bounded shear flows \citep{Farrell1993, Bamieh2001,Jovanovic2005,McKeon2010,mckeon2017engine}. This analysis framework has advantages of computational tractability and is not subject to finite channel effects. Related analysis has shown promise in studying stratified flows including inviscid stratified shear flow with constant shear \citep{farrell1993transient}, stratified PCF \citep{jose2015analytical,jose2018optimal}, and stratified turbulent channel flow \citep{ahmed2021resolvent}.

In this work, we extend the \emph{structured} input--output analysis (SIOA) originally developed for unstratified PCF \citep{liu2021structuredJournal} to stratified PCF. Prior application of the SIOA approach to unstratified transitional wall-bounded shear flows \citep{liu2021structuredJournal} demonstrated that including the componentwise structure of nonlinearity  uncovers a wider range of known key flow features identified through nonlinear analysis, experiments, and DNS,  but not captured through traditional (unstructured) input-output approaches. Here, SIOA for stratified PCF includes the effect of  nonlinearity in the momentum and density equations (under the Boussinesq approximation) within a computationally tractable linear framework through a feedback interconnection between the linearized dynamics and a structured forcing that is explicitly constrained to preserve the componentwise structure of the nonlinearity. The structured singular value \citep{doyle1982analysis,safonov1982stability} of the spatio-temporal frequency response associated with this feedback interconnection can then be calculated at each streamwise and spanwise length scale. This structured singular value can be interpreted as the flow structure that shows the largest input--output gain (amplification) given the structured feedback interconnection. 

Here, we apply the SIOA to characterize highly amplified flow structures in the intermittent regime of stratified PCF and investigate the behavior of the flow across a range of $Re$, $Ri_b$ and $Pr$. Our aims are two-fold. First, we wish to investigate whether the structures
predicted by the SIOA can be quantitatively identified with fully nonlinear structures that have been observed in previously reported DNS of stratified PCF with specific values of the control parameters $Re$, $Ri_b$ and $Pr$. Second, we wish to explore the dependence on the control parameters of predictions from the SIOA in parameter regimes which are not (as yet) accessible to DNS. More specifically, to address our first aim, we examine how $Re$ and $Ri_b$ affect flow structures with Prandtl number set at $Pr=0.7$, i.e. the value for air. We demonstrate that SIOA does indeed predict the characteristic wavelengths and angle of the oblique turbulent bands observed in very large channel size DNS of the low-$Re$ low-$Ri_b$ intermittent regime of stratified PCF at the same $Pr$ \citep{deusebio2015intermittency,taylor2016new}. We further show that in the high-$Re$ high-$Ri_b$ intermittent regime, the SIOA  identifies quasi-horizontal flow structures resembling turbulent-laminar layers \citep{deusebio2015intermittency}. 

Having achieved our first aim, and demonstrated the usefulness of 
the SIOA for identifying realistic nonlinear flow structures, we then turn our attention to our second aim. We demonstrate that increasing bulk Richardson number reduces the amplification of streamwise varying flow structures. These results show that the classical Miles-Howard stability criterion ($Ri_b\leq 1/4$) appears (perhaps fortuitously) to be associated with a change in the most amplified flow structures, which is robust for a wide range of $Re$ and valid at $Pr\approx 1$.

We then examine flow behavior at different $Ri_b$ and $Pr$. For flows with $Pr\ll 1$, a larger bulk Richardson number is required to reduce the amplification of streamwise varying flow structures to the same level as streamwise independent ones compared with $Pr\approx 1$. The largest amplification also is predicted to occur at the same value of the product $Pr Ri_b$ consistent with the observation of the averaged density profile only varying with the product $Pr Ri_b$ in the $Pr\ll 1$ regime  \citep{langham2020stably}. For flows with $Pr\gg 1$, the SIOA identifies another quasi-horizontal flow structure independent of $Ri_b$. By decomposing input--output pathways into separate components associated with velocity and density fluctuations, we show that these quasi-horizontal flow structures at $Pr\gg 1$ are primarily associated with fluctuations in the density field. We further highlight the importance of this density-associated flow structure at $Pr\gg 1$ by constructing an analytical scaling argument for the input-output amplification in terms of $Re$ and $Pr$ under the assumptions of unstratified flow (with $Ri_b=0$) and streamwise invariance. The above observations using SIOA distinguish two types of quasi-horizontal flow structures, one associated with the high-$Re$ high-$Ri_b$ regime and the other one associated with density perturbations that emerges in the high $Pr$ regime.

To achieve our twin aims, and to demonstrate the above summarised results, the remainder of this paper is organized as follows. Section \ref{sec:stratified_structured_IO} describes the flow configuration of stratified plane Couette flow and then develops the structured input--output analysis for this flow. Section \ref{sec:stratified_result} analyzes the results obtained from structured input--output analysis focusing on the wall-parallel length scale of flow structures in this flow. 
In \S\ \ref{sec:stratified_Re_Pr_scaling_passive_limit}, we  develop analytical scaling arguments with respect to  $Re$ and $Pr$ to investigate behavior for flows in the high $Pr$ limit. Finally we draw conclusions and suggest some avenues of future work in \S\ \ref{sec:stratified_stratified_conclusion}.

\section{Structured input--output response of stratified flow}
\label{sec:stratified_structured_IO}

\subsection{Governing Equations}

\begin{figure}
(a) \hspace{0.5\textwidth} (b)

\centering
\includegraphics[width=0.49\textwidth]{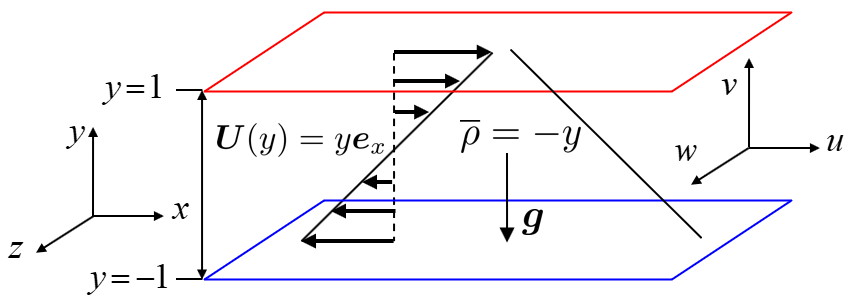}
\includegraphics[width=0.49\textwidth]{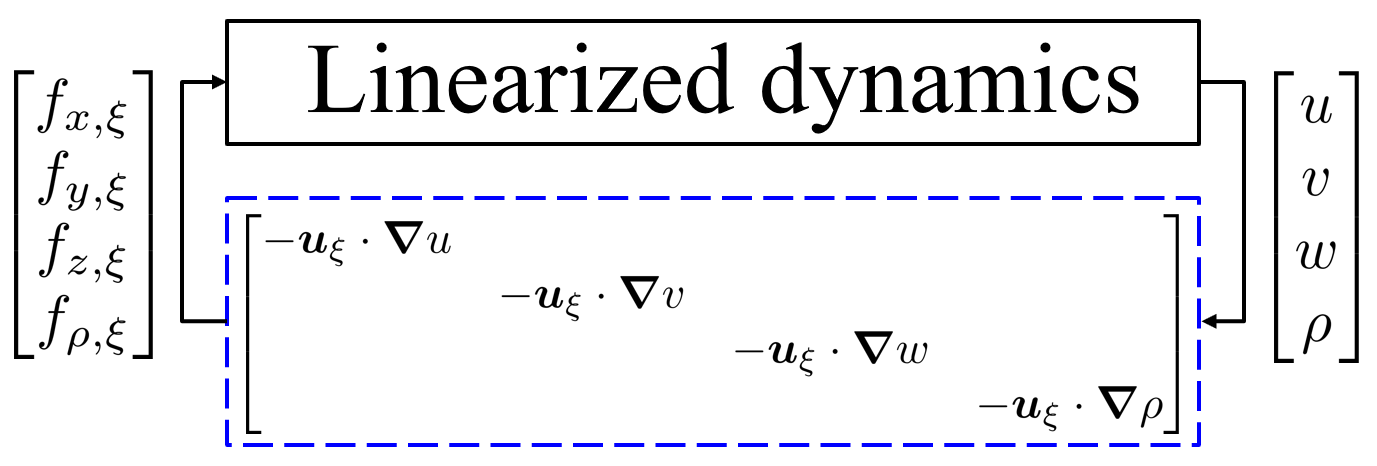}
    \caption{(a) Schematic of stably stratified plane Couette flow with laminar base flow $\boldsymbol{U}(y)=y\boldsymbol{e}_x$ and background density $\overline{\rho}=-y$. The gravity $\boldsymbol{g}=-g\boldsymbol{e}_y$ force is orthogonal to channel walls. The blue and red colors represent high-density and low-density fluids, respectively. (b) Block diagram of the feedback interconnection between the linearized dynamics and structured forcing (outlined by the blue dashed line) in \eqref{eq:stratified_f_all_uncertain_model} modeling the nonlinearity. }
    \label{fig:stratified_flow_config_cou_poi}
\end{figure}

We consider stably stratified plane Couette flow (PCF) between two infinite parallel plates and employ $x$, $y$, and $z$ to denote the streamwise, wall-normal (or vertical), and spanwise directions. The corresponding (assumed incompressible) velocity components are denoted as $u$, $v$, and $w$. The coordinate frames and configurations for this stratified PCF are shown in figure \ref{fig:stratified_flow_config_cou_poi}. We express the velocity field as a vector $\boldsymbol{u}_{\text{tot}}=\begin{bmatrix}u_{\text{tot}} & v_{\text{tot}} & w_{\text{tot}}\end{bmatrix}^\text{T}$ with $^\text{T}$ indicating the transpose. We then decompose the velocity field into the sum of a laminar  linearly-varying base flow $U(y)=y$ and fluctuations about the base flow $\boldsymbol{u}$; i.e., $\boldsymbol{u}_{\text{tot}}=U(y)\boldsymbol{e}_x+\boldsymbol{u}$ with $\boldsymbol{e}_x$ denoting the $x$-direction (streamwise) unit vector. The pressure field is similarly decomposed as $p_{\text{tot}}=P+p$. We decompose the density $\rho_{\text{tot}}$ as the sum of a reference density $\rho_r$, a base, again linearly-varying, density $\overline{\rho}=-y$ and a density fluctuation $\rho$; i.e., $\rho_{\text{tot}}=\rho_r+\overline{\rho}+\rho$. We use $\rho_0$ to denote half of the density difference between the top and bottom walls, and it is assumed to be much smaller than the reference density $ \rho_0\ll\rho_r$ so that the Boussinesq approximation can be used. 

The dynamics of the fluctuations $\boldsymbol{u}$, $p$, and $\rho$ are hence governed by the Navier-Stokes equations for an incompressible velocity field under the Boussinesq approximation and an advection-diffusion equation for the density:
\begin{subequations} \label{eq:stratified_NS_All}
\begin{align}
\partial_{t} \boldsymbol{u}  
+  U\partial_x \boldsymbol{u}  + v\,\frac{dU }{dy}\boldsymbol{e}_x+Ri_b\;\rho\,\boldsymbol{e}_y +\boldsymbol{\nabla} p
-\frac{1}{Re}{\nabla}^2 \boldsymbol{u}
 &=- \boldsymbol{u} {\cdot} \boldsymbol{\nabla} \boldsymbol{u}, \label{eq:stratified_NSDecompf_momentum} \\
 \partial_t \rho+U\partial_x \rho+v\frac{d\overline{\rho}}{dy}-\frac{1}{RePr}\nabla^2 \rho &=-\boldsymbol{u}{\cdot}  \boldsymbol{\nabla}\rho ,\label{eq:stratified_NSDecompf_density}\\
\boldsymbol{\nabla} {\cdot} \boldsymbol{u}&=0. \label{eq:stratified_NSDecompf_divergence_free}
\end{align}
\end{subequations} 
Here, the spatial variables are normalized by the channel half-height $h$, the velocity is normalized by half of the velocity difference between the top and bottom walls $U_w$, where $\pm U_w$ is the velocity at the 
channel walls. Time and pressure are normalized by $h/U_w$ and $\rho_r U_w^2$, respectively. The base density field $\overline{\rho}(y)$ and the density fluctuations $\rho$ are normalized by $\rho_0$. Under this normalization, the base density profile $\overline{\rho}=-y$ is balanced by a hydrostatic pressure $P=Ri_b y^2/2$. 

The nondimensional control parameters are the Reynolds number $Re$, the Prandtl number $Pr$, and the bulk Richardson number $Ri_b$, naturally defined as:
\begin{subequations}
\begin{align}
    Re:=\frac{U_wh}{\nu},\;\;
    Pr:=\frac{\nu}{\kappa},\;\;Ri_b:=\frac{g \rho_0 h}{\rho_r U_w^2},\tag{\theequation a-c}
\end{align}
\end{subequations}
where $\nu$ is the kinematic viscosity, $\kappa$ is the molecular diffusivity of the density scalar and $g$ is the magnitude of gravity. The gravity is in the direction orthogonal to the wall $\boldsymbol{g}=-g\boldsymbol{e}_y$ with $\boldsymbol{e}_y$ denoting the $y$-direction (wall-normal, or vertical) unit vector. In equation set \eqref{eq:stratified_NS_All}, $\boldsymbol{\nabla}:=\begin{bmatrix}\partial_x & \partial_y & \partial_z\end{bmatrix}^\text{T}$ represents the gradient operator, and $\nabla^2:=\partial_{x}^2+\partial_{y}^2+\partial_{z}^2$ represents the Laplacian operator. We impose no-slip boundary conditions at the wall $\boldsymbol{u}(y=\pm 1)=\boldsymbol{0}$ and Dirichlet boundary conditions for density fluctuations $\rho(y=\pm 1)=0$ that can be maintained by e.g., constant temperatures at the wall with a linear equation of state (with the hotter plate at the top).

A large body of linear analysis techniques views the nonlinear terms as a forcing, which enables these terms to be represented as an unmodeled effect (which can be thought of as some type of `uncertainty' in the equations).  There are a wide range of such models, but a common approach is a delta-correlated or colored stochastic forcing that captures a wide range of the unmodeled effects, see e.g., the discussion in \citet{Farrell1993, Bamieh2001,Jovanovic2005,McKeon2010,mckeon2017engine,Zare2017}. Here we similarly write the nonlinear terms as the forcing:
\begin{subequations}
\label{eq:stratified_f_nonlinear_all}
\begin{align}
\boldsymbol{f}_{\boldsymbol{u}}:=&-\boldsymbol{u}{\cdot} \boldsymbol{\nabla }\boldsymbol{u}=\begin{bmatrix}-\boldsymbol{u}{\cdot} \boldsymbol{\nabla }u &-\boldsymbol{u}{\cdot} \boldsymbol{\nabla }v& -\boldsymbol{u}{\cdot} \boldsymbol{\nabla }w\end{bmatrix}^\text{T}=:\begin{bmatrix}f_x&f_y&f_z\end{bmatrix}^\text{T}. 
     \label{eq:stratified_f_nonlinear}\\
     f_{\rho}:=&-\boldsymbol{u}{\cdot} \boldsymbol{\nabla }\rho,
    \label{eq:stratified_f_nonlinear_density}
\end{align}
\end{subequations}
which turns \eqref{eq:stratified_NS_All} into a set of linear evolution equations subject to the forcing terms $\boldsymbol{f}_{\boldsymbol{u}}$ and $f_{\rho}$.

We now construct the model of the nonlinearity, where the velocity field $-\boldsymbol{u}$ in \eqref{eq:stratified_f_nonlinear_all} associated with the forcing components can be viewed as the gain operator of an input--output system in which the velocity and density gradients  $\boldsymbol{\nabla} u$, $\boldsymbol{\nabla} v$, $\boldsymbol{\nabla} w$, $\boldsymbol{\nabla}\rho$ act as the respective inputs and the forcing components $f_x$, $f_y$, $f_z$, and $f_{\rho}$ act as the respective output. This input--output model of the nonlinear components in the momentum and density equations, \eqref{eq:stratified_f_nonlinear_all}, are respectively given by
\begin{subequations}
\label{eq:stratified_f_all_uncertain_model}
\begin{align}
    \boldsymbol{f}_{\boldsymbol{u},\xi}:=&-\boldsymbol{u}_{\xi}{\cdot} \boldsymbol{\nabla }\boldsymbol{u}=\begin{bmatrix}-\boldsymbol{u}_{\xi}{\cdot} \boldsymbol{\nabla }u &-\boldsymbol{u}_{\xi}{\cdot} \boldsymbol{\nabla }v& -\boldsymbol{u}_{\xi}{\cdot} \boldsymbol{\nabla }w\end{bmatrix}^\text{T}=:\begin{bmatrix}f_{x,\xi}&f_{y,\xi}&f_{z,\xi}\end{bmatrix}^\text{T},
     \label{eq:stratified_f_uncertain_model}\\
     f_{\rho,\xi}:=&-\boldsymbol{u}_{\xi}{\cdot} \boldsymbol{\nabla}\rho. 
    \label{eq:stratified_f_density_uncertain_model}
\end{align}
\end{subequations}
Here, $-\boldsymbol{u}_{\xi}$ in equations \eqref{eq:stratified_f_all_uncertain_model}  maps the corresponding velocity and density gradient into each component of the modeled forcing driving linearized dynamics. This forcing in \eqref{eq:stratified_f_all_uncertain_model} is referred to as structured forcing because it preserves the componentwise structure of nonlinearity in  \eqref{eq:stratified_f_nonlinear_all}. Figure \ref{fig:stratified_flow_config_cou_poi}(b) shows a block diagram of the feedback interconnection between the linearized dynamics and this forcing whose block diagonal structure  mirrors the nonlinear interactions in the Navier Stokes equations, i.e. the forcing does not include terms such as $-\boldsymbol{u}\cdot\boldsymbol{\nabla} v$ in the forcing $f_{x,\xi}$ since this term does not appear in the Navier Stokes equations.

Although the nonlinearity in \eqref{eq:stratified_f_nonlinear_all} can be written in many different ways, the current formulation leads to a straightforward and unified formulation for structured forcing in each momentum and density equation in \eqref{eq:stratified_f_all_uncertain_model}. We next exploit this form of the equations to construct an input--output map using the structured singular value formalism \citep{packard1993complex,zhou1996robust}. This map will enable us to analyze the {fluctuations} which are prominent in the intermittent regime.

\subsection{Structured input--output response}
\label{subsec:stratified_structured_uncertainty}

We need to define the spatio-temporal frequency response $\mathcal{H}_\nabla^S(y;k_x,k_z,\omega)$ of stratified PCF that will form the basis of the structured input--output response. We use the superscript $^S$ to distinguish this operator from its counterpart for unstratified wall-bounded shear flow \citep{liu2021structuredJournal}. We employ the standard transformation to express the velocity field dynamics in \eqref{eq:stratified_NS_All} in terms of a formulation using the wall-normal velocity $v$ and the wall-normal vorticity $\omega_y:=\partial_z u-\partial_x w$ \citep{schmid2012stability}. This transformation enforces the incompressibility constraint in \eqref{eq:stratified_NSDecompf_divergence_free} and eliminates the pressure by construction. We  exploit shift-invariance in the $(x,z)$ spatial directions and assume shift-invariance in time $t$, which allows us to perform the following triple Fourier transform:\begin{equation}
\widehat{\psi}(y;k_x,k_z,\omega):=\int\limits_{-\infty}^{\infty} \int\limits_{-\infty}^{\infty}\int\limits_{-\infty}^{\infty}\psi(x,y,z,t)e^{-\text{i}(k_x x + k_z z+\omega t )}\,dx\,dz\,dt,\label{eq:ftdef}
\end{equation} where $\text{i}=\sqrt{-1}$, $\omega$ is the temporal frequency, and $k_x = 2\pi/\lambda_x$ and $k_z = 2\pi/\lambda_z$ are the $x$ and $z$ wavenumbers, respectively. The sign of temporal frequency $\omega$ in \eqref{eq:ftdef} is chosen for ease of employing control-oriented toolboxes.

The resulting equations describing the transformed linearized equations subject to the forcing $\begin{bmatrix}\boldsymbol{f}_{\boldsymbol{u}, \xi}\\ f_{\rho,\xi}\end{bmatrix}$ are given by
\begin{subequations}
\label{eq:stratified_ABC_frequency}
\begin{align}
    \text{i}\omega\begin{bmatrix}
    \widehat{v}\\
    \widehat{\omega}_y\\
    \widehat{\rho}
    \end{bmatrix}=&\;\widehat{\mathcal{A}}^S\begin{bmatrix}
    \widehat{v}\\
    \widehat{\omega}_y\\
    \widehat{\rho}
    \end{bmatrix}+\widehat{\mathcal{B}}^S\begin{bmatrix}\widehat{f}_{x,\xi}\\
    \widehat{f}_{y,\xi}\\
    \widehat{f}_{z,\xi}\\\widehat{f}_{\rho,\xi}\end{bmatrix},\\
    \begin{bmatrix}\widehat{u}\\
    \widehat{v}\\
    \widehat{w}\\
    \widehat{\rho}\end{bmatrix}=&\;\widehat{\mathcal{C}}^S\begin{bmatrix}
    \widehat{v}\\
    \widehat{\omega}_y\\
    \widehat{\rho}
    \end{bmatrix}.
\end{align}
\end{subequations}
The operators in equation set \eqref{eq:stratified_ABC_frequency} are defined as:
\begingroup
\allowdisplaybreaks
\begin{subequations}
\label{eq:stratified_operator_ABC}
\begin{align}
    \widehat{\mathcal{A}}^S(k_x,k_z):=&\;\widehat{\mathcal{M}}^{-1} \begin{bmatrix}
    -\text{i}k_xU{\widehat{\nabla}}^2+\text{i}k_xU''+\frac{\widehat{{\nabla}}^4}{Re} & 0 & Ri_b(k_x^2+k_z^2)\\
    -\text{i}k_zU' & -\text{i}k_x U+\frac{\widehat{{\nabla}}^2}{Re} & 0\\
    -\overline{\rho}' & 0 & -\text{i}k_x U+\frac{\widehat{\nabla}^2}{Re Pr}
    \end{bmatrix},\label{eq:stratified_operator_ABC_a}\\
    \mathcal{\widehat{B}}^S(k_x,k_z):=&\;\widehat{\mathcal{M}}^{-1}
    \begin{bmatrix}
    -\text{i}k_x\partial_y & -(k_x^2+k_z^2) & -\text{i}k_z \partial_y & 0\\
    \text{i}k_z & 0 & -\text{i}k_x & 0\\
    0 & 0 & 0 & \mathcal{I}
    \end{bmatrix},\;\;\widehat{\mathcal{M}}:=\begin{bmatrix}\widehat{{\nabla}}^2 & 0 & 0\\
    0 & \mathcal{I} & 0\\
    0 & 0 & \mathcal{I}
    \end{bmatrix},\label{eq:stratified_operator_ABC_b}\\
    \mathcal{\widehat{C}}^S(k_x,k_z):=&\;\frac{1}{k_x^2+k_z^2}\begin{bmatrix}
    \text{i}k_x\partial_y & -\text{i}k_z & 0 \\
    k_x^2+k_z^2 & 0 & 0\\
    \text{i}k_z \partial_y & \text{i}k_x & 0 \\
    0 & 0 & k_x^2+k_z^2
    \end{bmatrix},\label{eq:stratified_operator_ABC_c}
\end{align}
\end{subequations}
\endgroup
where $U':=dU(y)/dy$, $U'':=d^2U(y)/dy^2$, $\overline{\rho}':=d\overline{\rho}(y)/dy$, $\widehat{{\nabla}}^2:=\partial_{yy}-k_x^2-k_z^2$, $\widehat{\nabla}^4:=\partial^{(4)}_{y}-2(k_x^2+k_z^2)\partial_{yy}+(k_x^2+k_z^2)^2$, and $\mathcal{I}$ is the identity operator. The equation associated with $\widehat{\mathcal{A}}^S$ operator in \eqref{eq:stratified_operator_ABC_a} can also be obtained by generalizing  the classical Taylor-Goldstein equation \citep{taylor1931effect,goldstein1931stability,smyth2019instability} to include viscosity, density diffusivity, and coupling with wall-normal vorticity $\widehat{\omega}_y$ with $k_z\neq 0$. The boundary conditions associated with \eqref{eq:stratified_ABC_frequency} are
$\widehat{v}(y=\pm 1)=
    \frac{\partial \widehat{v}}{\partial y}(y=\pm1)=\widehat{\omega}_y(y=\pm 1)=\widehat{\rho}(y=\pm 1)=0$.

The spatio-temporal frequency response $\mathcal{H}^S$ of the system in \eqref{eq:stratified_ABC_frequency}, which maps the input forcing to the velocity and density fields at the same spatial-temporal wavenumber-frequency triplet; i.e., $\begin{bmatrix}\boldsymbol{\widehat{u}}(y;k_x,k_z,\omega)\\{\widehat{\rho}}(y;k_x,k_z,\omega)
\end{bmatrix}=\mathcal{H}^S(y;k_x,k_z,\omega)\begin{bmatrix}\boldsymbol{\widehat{f}}_{\boldsymbol{u}, \xi}(y;k_x,k_z,\omega)\\
\widehat{f}_{\rho,\xi}(y;k_x,k_z,\omega)\end{bmatrix}$, is given by 
\begin{equation}
    \mathcal{H}^S(y;k_x,k_z,\omega):=\widehat{\mathcal{C}}^S\left(\text{i}\omega\,\mathcal{I}_{3\times 3}-\widehat{\mathcal{A}}^S\right)^{-1}\widehat{\mathcal{B}}^S.
     \label{eq:stratified_linearized_ABC}
\end{equation}
Here $\mathcal{I}_{3\times 3}:=\text{diag}(\mathcal{I},\mathcal{I},\mathcal{I})$, where $\text{diag}(\cdot)$ indicates a block diagonal operation. 

\begin{figure}

	(a) \hspace{0.7\textwidth} (b)
	
    \centering
    \includegraphics[width=0.7\textwidth]{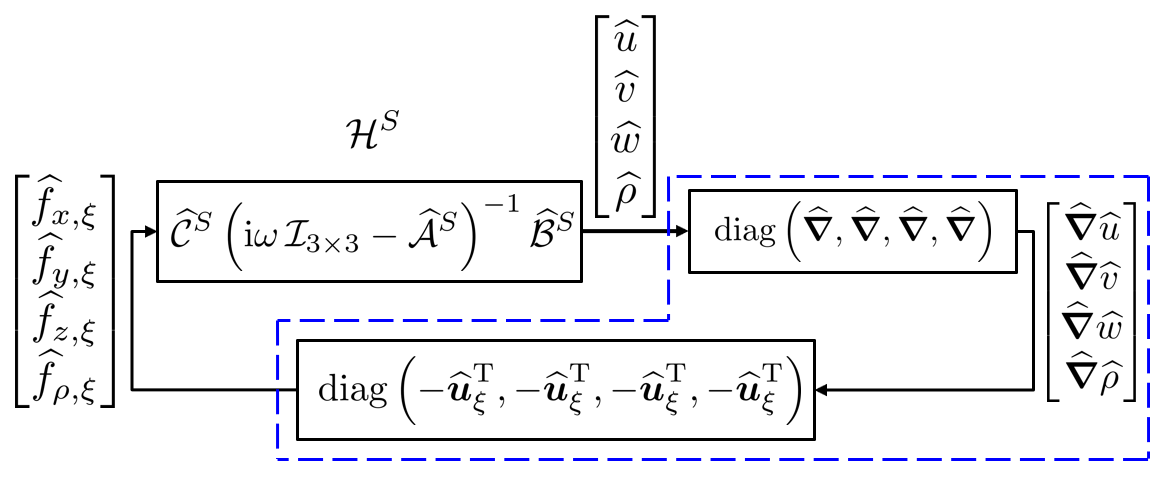}
    \hspace{0.2in}
    \includegraphics[width=0.8in,trim=-0 -0.5in 0 0]{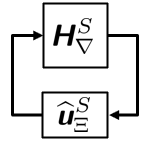}
    \caption{  Block diagram showing the feedback interconnection of the structured input--output analysis (SIOA) framework applied to stratified plane Couette flow (PCF). Panel (a) redraws figure \ref{fig:stratified_flow_config_cou_poi}(b),  where the blocks inside of ({\color{blue}$\dashed$}, blue) lines represent the modeled forcing in equation \eqref{eq:stratified_feedback_structured_uncertainty} corresponding to the bottom block in figure \ref{fig:stratified_flow_config_cou_poi}(b) also inside of ({\color{blue}$\dashed$}, blue). Panel (b) redraws panel (a) after discretization with the top block corresponding to the combination of the two top blocks in panel (a) and the bottom block corresponding to the bottom block of the panel (a).}
    \label{fig:stratified_feedback_detail}
\end{figure}

The linear form of \eqref{eq:stratified_f_uncertain_model}-\eqref{eq:stratified_f_density_uncertain_model}  allows us to perform the same spatio-temporal Fourier transform on the model of the nonlinearity, which can be decomposed as
\begin{equation}
\begin{bmatrix}
    \widehat{f}_{x,\xi}\\
    \widehat{f}_{y,\xi}\\
    \widehat{f}_{z,\xi}\\
    \widehat{f}_{\rho,\xi}
    \end{bmatrix}=\text{diag}\left(-\widehat{\boldsymbol{u}}_{\xi}^{\text{T}} ,-\widehat{\boldsymbol{u}}_{\xi}^{\text{T}},-\widehat{\boldsymbol{u}}_{\xi}^{\text{T}}, -\widehat{\boldsymbol{u}}_{\xi}^{\text{T}} \right)\text{diag}\left(\boldsymbol{\widehat{\nabla}},\boldsymbol{\widehat{\nabla}},\boldsymbol{\widehat{\nabla}},\boldsymbol{\widehat{\nabla}}\right)\begin{bmatrix}
\widehat{u}\\
\widehat{v}\\
\widehat{w}\\
\widehat{\rho}
\end{bmatrix}.
\label{eq:stratified_feedback_structured_uncertainty}
\end{equation}
A block diagram illustrating this decomposition of the nonlinearity is shown inside the blue dashed line ({\color{blue}$\dashed$}) in figure \ref{fig:stratified_feedback_detail}(a). This block diagonal structure constrains the modeled nonlinear interactions, i.e., provides structured forcing.

In order to isolate the gain operator
$-\boldsymbol{u}_{\xi}$, we combine the linear gradient operator with the spatio-temporal frequency response of the linearized system \eqref{eq:stratified_linearized_ABC}. The resulting modified frequency response operator with outputs that are the vectorized gradients of the velocity and density components is defined as
\begin{equation}
     \mathcal{H}^S_{\nabla}(y;k_x,k_z,\omega):=\text{diag}\left(\boldsymbol{\widehat{\nabla}},\boldsymbol{\widehat{\nabla}},\boldsymbol{\widehat{\nabla}},\boldsymbol{\widehat{\nabla}}\right)\mathcal{H}^S(y;k_x,k_z, \omega).
    \label{eq:stratified_H_operator_grad}
\end{equation}
The resulting system model can be redrawn as a feedback interconnection between this linear operator and the structured uncertainty
\begin{equation}
    \widehat{\boldsymbol{u}}^S_{\Upxi}:=\text{diag}\left(-\widehat{\boldsymbol{u}}_{\xi}^{\text{T}},-\widehat{\boldsymbol{u}}_{\xi}^{\text{T}} ,-\widehat{\boldsymbol{u}}_{\xi}^{\text{T}},-\widehat{\boldsymbol{u}}_{\xi}^{\text{T}} \right).
    \label{eq:stratified_Delta_block_diag}
\end{equation}
Here the structure is introduced in terms of the diagonal form of $\widehat{\boldsymbol{u}}_{\Upxi}^S$ that enforces the componentwise structure of the nonlinearity in the forcing model defined in \eqref{eq:stratified_f_all_uncertain_model}. Figure \ref{fig:stratified_feedback_detail}(b) illustrates this feedback interconnection between the modified spatio-temporal frequency response and the structured uncertainty, where $\mathsfbi{H}^S_{\nabla}$ and $\mathbfsbilow{\widehat{u}}^S_{\Upxi}$ respectively represent the spatial discretizations (numerical approximations) of $\mathcal{H}^S_{\nabla}$ in \eqref{eq:stratified_H_operator_grad} and $\boldsymbol{\widehat{u}}^S_{\Upxi}$ in \eqref{eq:stratified_Delta_block_diag}.

We are interested in characterizing the horizontal length scales of the most amplified flow structures under this structured forcing. This amplification  can be quantified in terms of the structured singular value of the modified frequency response operator $\mathcal{H}^S_\nabla$; see e.g., \citet[definition 3.1]{packard1993complex}; \citet[definition 11.1]{zhou1996robust}, which is defined as follows.

\begin{definition}\label{def:mu_stratified} Given wavenumber and frequency pair $(k_x,k_z,\omega)$, the structured singular value $\mu_{\mathbfsbilow{\widehat{U}}^S_{\Upxi}}\left[\mathbfsbilow{H}^S_{\nabla}(k_x,k_z,\omega)\right]$ is defined as:
\begin{align}
    \mu_{\mathbfsbilow{\widehat{U}}^S_{\Upxi}}\left[\mathbfsbilow{H}^S_{\nabla}(k_x,k_z,\omega)\right]:=\frac{1}{\min\{\bar{\sigma}[\mathbfsbilow{\widehat{u}}^S_{\Upxi}]\,:\,\mathbfsbilow{\widehat{u}}^S_{\Upxi}\in \mathbfsbilow{\widehat{U}}^S_{\Upxi},\,\det[\mathsfbi{I}-\mathbfsbilow{H}^S_{\nabla}(k_x,k_z,\omega)\mathbfsbilow{\widehat{u}}^S_{\Upxi}]=0\}}.\label{eq:stratified_mu}
\end{align}
If no $\mathbfsbilow{\widehat{u}}^S_{\Upxi}\in \mathbfsbilow{\widehat{U}}^S_{\Upxi}$ makes $\mathsfbi{I}-\mathbfsbilow{H}^S_{\nabla}\mathbfsbilow{\widehat{u}}^S_{\Upxi}$ singular, then $\mu_{\mathbfsbilow{\widehat{U}}^S_{\Upxi}}[\mathbfsbilow{H}^S_{\nabla}]:=0$.

Here, $\bar{\sigma}[\cdot]$ is the largest singular value, $\det[\cdot]$ is the determinant of the argument, and $\mathsfbi{I}$ is the identity matrix. The subscript of $\mu$ in \eqref{eq:stratified_mu} is a set $\mathbfsbilow{\widehat{U}}^S_{\Upxi}$ containing all uncertainties having the same block-diagonal structure as $\mathbfsbilow{\widehat{u}}^S_{\Upxi}$; i.e.,
\begin{align}
\mathbfsbilow{\widehat{U}}^S_{\Upxi}:=\left\{{\rm {diag}}\left(-\mathbfsbilow{\widehat{u}}_{\xi}^{\text{T}},-\mathbfsbilow{\widehat{u}}_{\xi}^{\text{T}},-\mathbfsbilow{\widehat{u}}_{\xi}^{\text{T}},-\mathbfsbilow{\widehat{u}}_{\xi}^{\text{T}}\right):-\mathbfsbilow{\widehat{u}}^{\text{T}}_{\xi}\in \mathbb{C}^{N_y\times 3N_y}\right\},
\label{eq:stratified_uncertain_set}
\end{align}
where $N_y$ denotes the number of grid points in $y$.
\end{definition}

Definition \ref{def:mu_stratified} suggests that the inverse of the structured singular value $1/\mu$ is the minimal norm of the perturbation $\mathbfsbilow{\widehat{u}}^S_{\Upxi}$ that destabilizes the feedback interconnection in figure \ref{fig:stratified_feedback_detail}(b) in the input-output sense defined by the small gain theorem, see \citet[proposition 2.2]{liu2021structuredJournal} and \citet[theorem 11.8]{zhou1996robust}. This interpretation suggests that the flow field is more sensitive to perturbations with the flow structures associated with a larger amplification measured by the structured singular value $\mu$. A similar notion of destabilizing perturbation was also employed to interpret the largest (unstructured) singular value \citep[p. 581]{Trefethen1993Hydrodynamic}.

Here, the form of structured uncertainty in (\ref{eq:stratified_uncertain_set}) allows the full degrees of freedom for the complex matrix $ -\mathbfsbilow{\widehat{u}}^{\text{T}}_{\xi}\in \mathbb{C}^{N_y\times 3N_y}$ for ease of computation. While $\boldsymbol{u}_\xi$ is not constrained to be incompressible,  the incompressibility of $\boldsymbol{u}$ and the role of pressure are accounted for within the current $v$-$\omega_y$ formulation. Further refinement to better represent the physics and uncover the form of $\boldsymbol{u}_\xi$ requires an extension of both the analysis method and computational tools. These extensions are beyond the scope of the current work.

We then define the structured response following \cite{liu2021structuredJournal} as:
\begin{align}
    \|\mathcal{H}^S_{\nabla}\|_{\mu}(k_x,k_z):=\underset{\omega \in \mathbb{R}}{\text{sup}}\,\mu_{\mathbfsbilow{\widehat{U}}^S_{\Upxi}}\left[\mathbfsbilow{H}^S_{\nabla}(k_x,k_z,\omega)\right],
    \label{eq:stratified_H_nabla_mu}
\end{align}
where $\text{sup}$ represents a supremum (least upper bound) operation. Here we abuse the notation and terminology by writing $\|\cdot\|_{\mu}$ (Packard \& Doyle, 1993), although this quantity is not a proper norm. We employ this notation in analogy with the corresponding unstructured response of the feedback interconnection, which is given by \begin{equation}
    \|\mathcal{H}^S_{\nabla}\|_{\infty}(k_x,k_z):=\underset{\omega \in \mathbb{R}}{\text{sup}}\,\bar{\sigma}\left[\mathbfsbilow{H}^S_{\nabla}(k_x,k_z,\omega)\right].\label{eq:stratified_H_nabla_inf}
\end{equation}
This quantity is the unstructured counterpart of $\|\mathcal{H}^S_{\nabla}\|_{\mu}$, which is obtained by replacing the structured uncertainty set $\mathbfsbilow{\widehat{U}}^S_{\Upxi}$ with the set of full complex matrices $\mathbb{C}^{4N_y\times 12N_y}$. In both cases, a larger value indicates that the corresponding flow structures (associated with a particular $k_x$ and $k_z$ pair) have larger amplification under either structured or unstructured feedback forcing. For example, a larger value of $\|\mathcal{H}^S_{\nabla}\|_{\mu}(k_x,k_z)$ indicates that the corresponding flow structures (associated with a particular $k_x$ and $k_z$ pair) have larger amplification under structured feedback forcing in figure \ref{fig:stratified_feedback_detail}(b).

\subsection{Numerical Method}
\label{subsec:stratified_numerical}

We employ the Chebyshev differential matrix \citep{Weideman2000,trefethen2000spectral} to discretize the operators in equation set \eqref{eq:stratified_operator_ABC}. Our code is validated through comparison with the unstratified plane Couette flow and Poiseuille flow results in \citet{jovanovic2004modeling,Jovanovic2005,schmid2007nonmodal}. The implementation of stratification is validated by reproducing the maximum growth rate of the linear normal mode in a layered stratified plane Couette flow determined by \citet[figures 3 and 6(a)]{eaves2017multiple}, as well as the linear stability predictions for the unstable stratification configuration in \citet[figure 1 and Appendix B]{olvera2017exact}. We use $N_y=60$ collocation points not including the boundary points over the wall-normal extent, as well as $48$ and $36$ logarithmically spaced streamwise and spanwise wavenumbers in the respective spectral ranges $k_x \in [10^{-4},10^{0.48}]$ and $k_z \in [10^{-2},10^{1.2}]$, unless otherwise mentioned. To verify that this resolution is sufficient to achieve grid convergence we recomputed selected results with 1.5 times the number of collocation points in the wall-normal direction and verified that the results did not change. The quantity $\|\mathcal{H}^S_{\nabla}\|_{\mu}$ in  \eqref{eq:stratified_H_nabla_mu} for each wavenumber pair $(k_x,k_z)$ is computed using the \texttt{mussv} command in the Robust Control Toolbox \citep{balas2005robust} of MATLAB. The arguments of \texttt{mussv} employed here include the state-space model of $\mathbfsbilow{H}^S_{\nabla}$ that samples the frequency domain adaptively. The \texttt{BlockStructure} argument comprises four full $N_y\times 3N_y$ complex matrices, and we use the \texttt{`Uf'} algorithm option.

\section{Structured spatio-temporal frequency response of stratified flow}
\label{sec:stratified_result}

In this section, we use the structured input--output analysis (SIOA) approach described in \S\ \ref{subsec:stratified_structured_uncertainty} to characterize the flow structures that are most amplified in stably stratified plane Couette flow (PCF).

\subsection{Low-$Re$ low-$Ri_b$ versus high-$Re$ high-$Ri_b$ intermittency}
\label{subsec:stratified_Ri_b_Re}

\begin{figure}

    \centering   
     \includegraphics[width=\linewidth]{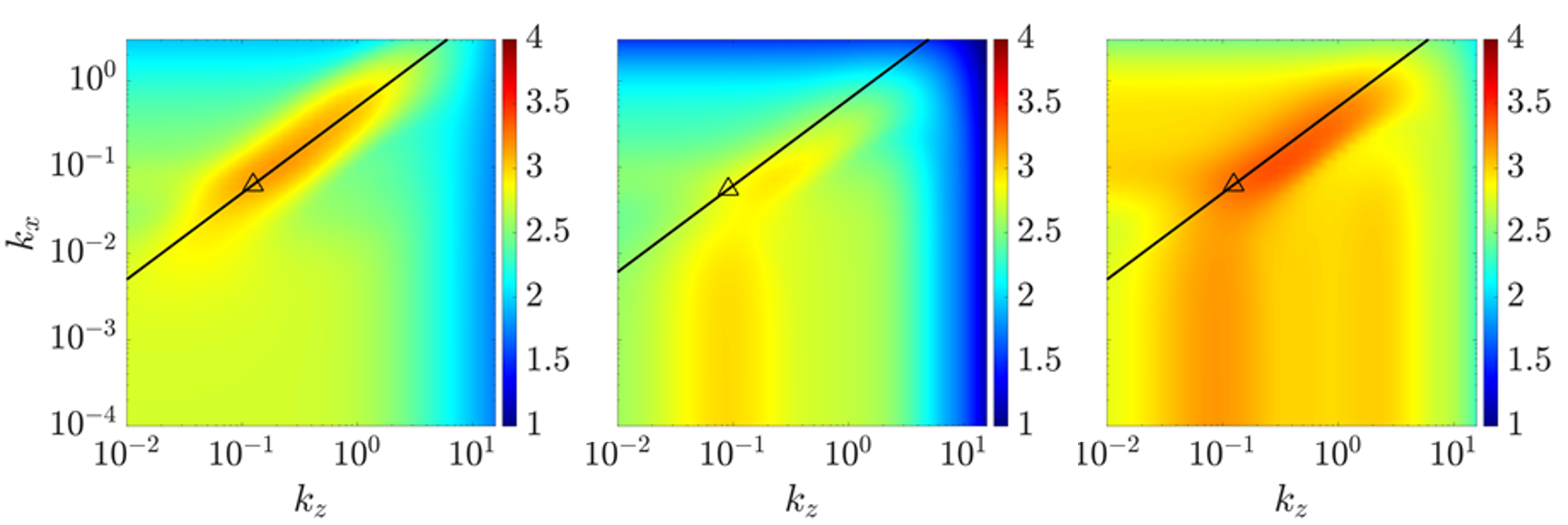}

    \caption{(a)$\text{log}_{10}[\|\mathcal{H}^S_{\nabla}\|_{\mu}(k_x,k_z)]$, (b)$\text{log}_{10}[\|\mathcal{H}^S\|_{\infty}(k_x,k_z)]$, and (c)$\text{log}_{10}[\|\mathcal{H}^S_{\nabla}\|_{\infty}(k_x,k_z)]$ for stratified plane Couette flow at $Re=865$, $Ri_b=0.02$, and $Pr=0.7$. Here the symbols ({\color{black}$\hspace{-0.0070in}\vspace{-0.015in}\triangle$}) are characteristic wavelengths ($\lambda_x=32\pi$, $\lambda_z=16\pi$) corresponding to the oblique turbulent band observed in DNS in the same flow regime \citep{deusebio2015intermittency,taylor2016new}. The lines ($\mline\mline$) are $\lambda_z=\lambda_x\tan(27^\circ)$ indicating the $27^\circ$  angle of the oblique turbulent bands.}
    \label{fig:stratified_mu_cou_stratified}
\end{figure}

In this subsection, we analyze flow structures that are prominent in either the low-$Re$ low-$Ri_b$ or the high-$Re$ high-$Ri_b$ intermittent regimes \citep{deusebio2015intermittency}. Here, we keep the Prandtl number fixed at $Pr=0.7$. This value  corresponds to thermally-stratified air and is the same value studied by \citet{deusebio2015intermittency}. We first consider a flow with $Re=865$, $Ri_b=0.02$ and $Pr=0.7$, where oblique turbulent bands have been observed \citep{deusebio2015intermittency,taylor2016new}. 
In order to evaluate the relative effect of the feedback interconnection and the imposed structure, we also compute $\|\mathcal{H}^S_{\nabla}\|_{\infty}(k_x,k_z)$ defined in \eqref{eq:stratified_H_nabla_inf} and
\begin{equation}
    \|\mathcal{H}^S\|_{\infty}(k_x,k_z):=\underset{\omega \in \mathbb{R}}{\text{sup}}\,\bar{\sigma}\left[\mathbfsbilow{H}^S(k_x,k_z,\omega)\right].\label{eq:stratified_H_inf}
    \end{equation}
Here, $\mathbfsbilow{H}^S$ is the discretization of spatio-temporal frequency response operator $\mathcal{H}^S$ in \eqref{eq:stratified_linearized_ABC}, i.e. the spatio-temporal frequency operator governing the linearized dynamics without the  feedback interconnection. The $\|\mathcal{H}\|_{\infty}$ for unstratified plane Couette and plane Poiseuille flows were  previously analyzed in \citet{jovanovic2004modeling,schmid2007nonmodal,illingworth2020streamwise}. The quantity in \eqref{eq:stratified_H_inf} describes the maximum singular value of the frequency response operator $\mathcal{H}^S$ which represents the maximal gain of $\mathcal{H}^S$ over all temporal frequencies; i.e., the worst-case amplification over harmonic inputs.

Figure \ref{fig:stratified_mu_cou_stratified} shows $\|\mathcal{H}^S_{\nabla}\|_{\mu}$ in panel (a) alongside (b) $\|\mathcal{H}^S\|_{\infty}$, and (c) $\|\mathcal{H}^S_{\nabla}\|_{\infty}$.
We indicate the characteristic wavelength pair $\lambda_x=32\pi$, $\lambda_z=16\pi$ corresponding to the oblique turbulent bands observed in DNS under the same flow regime \citep[figure 2(b)]{deusebio2015intermittency,taylor2016new} in these panels using the symbol ({\color{black}$\hspace{-0.0070in}\vspace{-0.015in}\triangle$}, black). These structures are observed to have a characteristic inclination angle (measured from the streamwise direction in $x-z$ plane) of $\theta:=\tan^{-1}(\lambda_z/\lambda_x)\approx27^\circ$, which is indicated in all panels by the black solid line ($\mline\mline$) that plots $\lambda_z=\lambda_x\tan(27^\circ)$. While there is some footprint of these structures and this angle in all three panels, the correspondence with the peak amplitude is most prominent in panel (a). In fact the peak value of $\|\mathcal{H}^S_{\nabla}\|_{\mu}$ in panel (a) occurs at 
streamwise and spanwise wavenumbers associated with the
characteristic wavelengths and angle of the oblique turbulent bands reported in DNS \citep{deusebio2015intermittency,taylor2016new}, and the line representing the angle of oblique turbulent bands crosses through the center of the narrow roughly elliptical peak region whose principle axis coincides with this angle. The results in figure \ref{fig:stratified_mu_cou_stratified}(a) suggest that the SIOA captures both the wavelengths and angle of the oblique turbulent bands in the low-$Re$ low-$Ri_b$ intermittent regime of stratified PCF. This analysis suggests that these oblique turbulent bands arise in the intermittent regime of stratified PCF due to their large amplification, or equivalently their sensitivity to disturbances. 

The traditional input--output analysis results, $\|\mathcal{H}^S\|_\infty$ in panel (b), provide a noticeable improvement compared with growth rate analysis (as presented in more detail in Appendix \ref{appendix:growth_rate}) and are also able to identify the preferred wavenumber pair in this intermittent regime. However, this analysis suggests larger amplification of the streamwise elongated modes. Moreover, the inclusion of an unstructured feedback loop quantified through  $\|\mathcal{H}^S_{\nabla}\|_{\infty}$ in panel (c)  correctly orders the relative amplification between the oblique turbulent bands and streamwise elongated structures ($k_x\approx 0$). The differences between $\|\mathcal{H}^S_{\nabla}\|_{\infty}$ and $\|\mathcal{H}^S\|_{\infty}$ are likely  associated with the additional $\widehat{\boldsymbol{\nabla}}$ operator in defining $\mathcal{H}_{\nabla}$ in \eqref{eq:stratified_H_operator_grad}, which emphasizes flow structures with a larger horizontal wavenumber. The difference between $\|\mathcal{H}^S_{\nabla}\|_{\mu}$ and $\|\mathcal{H}^S_{\nabla}\|_{\infty}$ is associated with the structured feedback interconnection that constrains the permissible feedback pathway, which weakens the amplification associated with the lift-up mechanism; see similar discussion on unstratified PCF \citep[section 3.3]{liu2021structuredJournal}. A comparison of the results in figure \ref{fig:stratified_mu_cou_stratified} suggests that it is the imposition of the componentwise structure from the nonlinear terms in \eqref{eq:stratified_f_nonlinear_all} further improves the prediction of the oblique turbulent bands.

\begin{figure}

	\hspace{0.05\textwidth}(a) \hspace{0.28\textwidth} (b) \hspace{0.28\textwidth} (c)
	
    \centering
    \includegraphics[width=\textwidth]{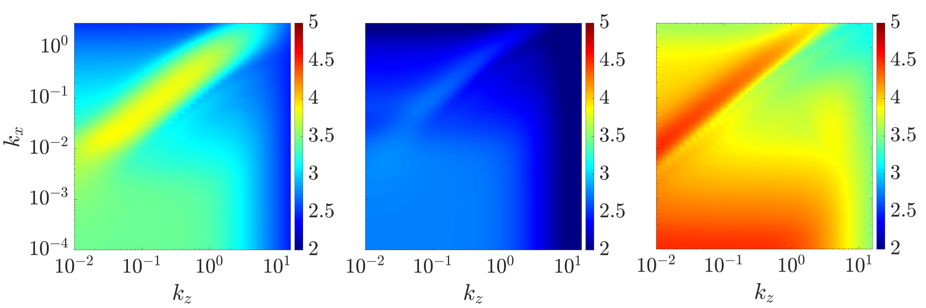}

    \caption{ $\text{log}_{10}[\|\mathcal{H}^S_{\nabla}\|_{\mu}(k_x,k_z)]$ at $Pr=0.7$ and: (a) $Re=4250$ and $Ri_b=0.02$; (b) $Re=865$ and $Ri_b=0.2541$, and (c) $Re=52630$ and $Ri_b=0.15$.}
    \label{fig:stratified_mu_high_Re_Ri_b}
\end{figure}

We now consider the high-$Re$ high-$Ri_b$ intermittent regime, which was shown to be qualitatively different in behavior from the low-$Re$ low-$Ri_b$ intermittent regime \citep{deusebio2015intermittency}. We first isolate the effect of increasing either $Re$ or $Ri_b$. Figure \ref{fig:stratified_mu_high_Re_Ri_b}(a) presents $\|\mathcal{H}^S_{\nabla}\|_{\mu}$ for a flow with $Ri_b=0.02$ and $Re=4250$. The larger  colorbar range versus figure \ref{fig:stratified_mu_cou_stratified} highlights the   expected higher magnitudes versus those for a flow with a lower Reynolds number ($Re=865$). We can see that the wavenumber pair of the peak region  extends towards smaller values (larger wavelengths) than those associated with the oblique turbulent bands that were in the peak region in figure \ref{fig:stratified_mu_cou_stratified}(a). Figure \ref{fig:stratified_mu_high_Re_Ri_b}(b) presents $\|\mathcal{H}^S_{\nabla}\|_{\mu}$ for a higher bulk Richardson number $Ri_b=0.2541$ and the same $Re$ and $Pr$ values as figure \ref{fig:stratified_mu_cou_stratified}(a). Here, the amplification associated with the streamwise varying flow structures such as the oblique turbulent bands observed in figure \ref{fig:stratified_mu_cou_stratified}(a) is reduced and  quasi-horizontal flow structures $(k_x\approx 0, k_z\approx 0)$ show a similar level of amplification (see the bottom left corner in figure \ref{fig:stratified_mu_high_Re_Ri_b}(b)). This flow structure associated with $k_x\approx 0$, $k_z\approx 0$ is referred to as quasi-horizontal to distinguish it from a horizontally uniform mode ($k_x=0$, $k_z=0$).

Armed with these insights, we consider the combined high-$Re$ high-$Ri_b$ intermittent regime ($Re=52630$ and $Ri_b=0.15$), which are the values corresponding to results shown in figure 7 of  \citet{deusebio2015intermittency}. Figure \ref{fig:stratified_mu_high_Re_Ri_b}(c) presents $\|\mathcal{H}^S_{\nabla}\|_{\mu}$ for these parameter values with an increased wall-normal grid with $N_y=90$. Here, the amplification of the oblique turbulent band is of a similar order to that of flow structures with a wide range of wavenumber pairs ranging from $k_x\lesssim 10^{-2}$ and $k_z\lesssim 1$ down to $k_x\approx 0$ and $k_z\approx 0$. These latter flow structures resemble the quasi-horizontal flow structures that have a horizontal length scale much larger than the vertical length scales, which are limited by the channel height and therefore restricted to nondimensional scales on the order of $2$.  The response in this regime, therefore, shows a large qualitative difference from that in the low-$Re$ low-$Ri_b$ ($Re=865$ and $Ri_b=0.02$) intermittent regime shown in figure \ref{fig:stratified_mu_cou_stratified}(a). This qualitative difference mirrors the different features in intermittent regimes described by \citet{deusebio2015intermittency}, where oblique turbulent bands are prevalent in the low-$Re$ low-$Ri_b$ intermittent regime, but the high-$Re$ high-$Ri_b$ intermittent regime is characterized by turbulent-laminar layers indicating a large horizontal length scale.

In figure \ref{fig:stratified_mu_high_Re_Ri_b}(c), we also observe that the quasi-horizontal flow structures have a streamwise wavelength much larger than their spanwise wavelength ($\lambda_x\gg \lambda_z$), which is also consistent with the observation in \citet{deusebio2015intermittency} that the turbulent and laminar layers in the high-$Re$ high-$Ri_b$ intermittent regime are homogeneous in the streamwise direction. This behavior can be understood through an order of magnitude estimation of the terms in the continuity equation. We assume highly anisotropic flow with $v\approx 0$ under strong stratification, which simplifies the continuity equation to $\frac{\partial u}{\partial x}+\frac{\partial w}{\partial z}=0$. We further assume that the restoring buoyancy force due to stratification does not have a preference between streamwise or spanwise directions and, therefore, we also assume $\frac{\partial u}{\partial x}$ and $\frac{\partial w}{\partial z}$ are the same order of magnitude. However, in the current stratified PCF configuration, streamwise velocity is associated with a characteristic velocity much larger than its spanwise counterpart due to the base flow velocity. As a result, streamwise variation is reduced much faster than spanwise variation (i.e., $k_x\ll k_z$) to keep $\frac{\partial u}{\partial x}$ and $\frac{\partial w}{\partial z}$  the same order of magnitude.

\subsection{$Ri_b>1/4$: a change in the most amplified flow structures}
\label{subsec:stratified_Ri_b_1_4_miles_howard}
 
\begin{figure}

	(a) \hspace{0.49\textwidth} (b)

    \centering
    \includegraphics[width=0.49\textwidth]{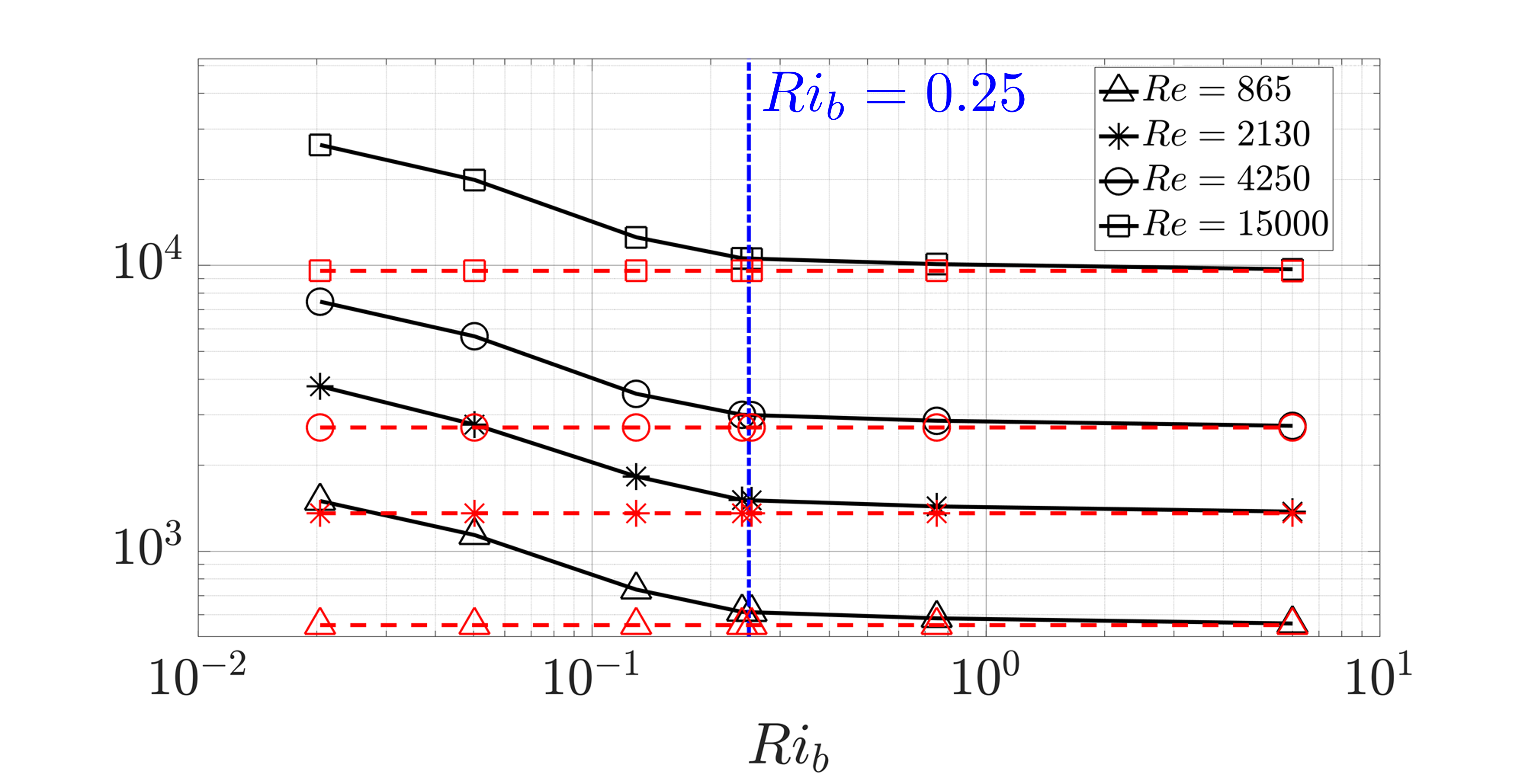}
    \includegraphics[width=0.49\textwidth]{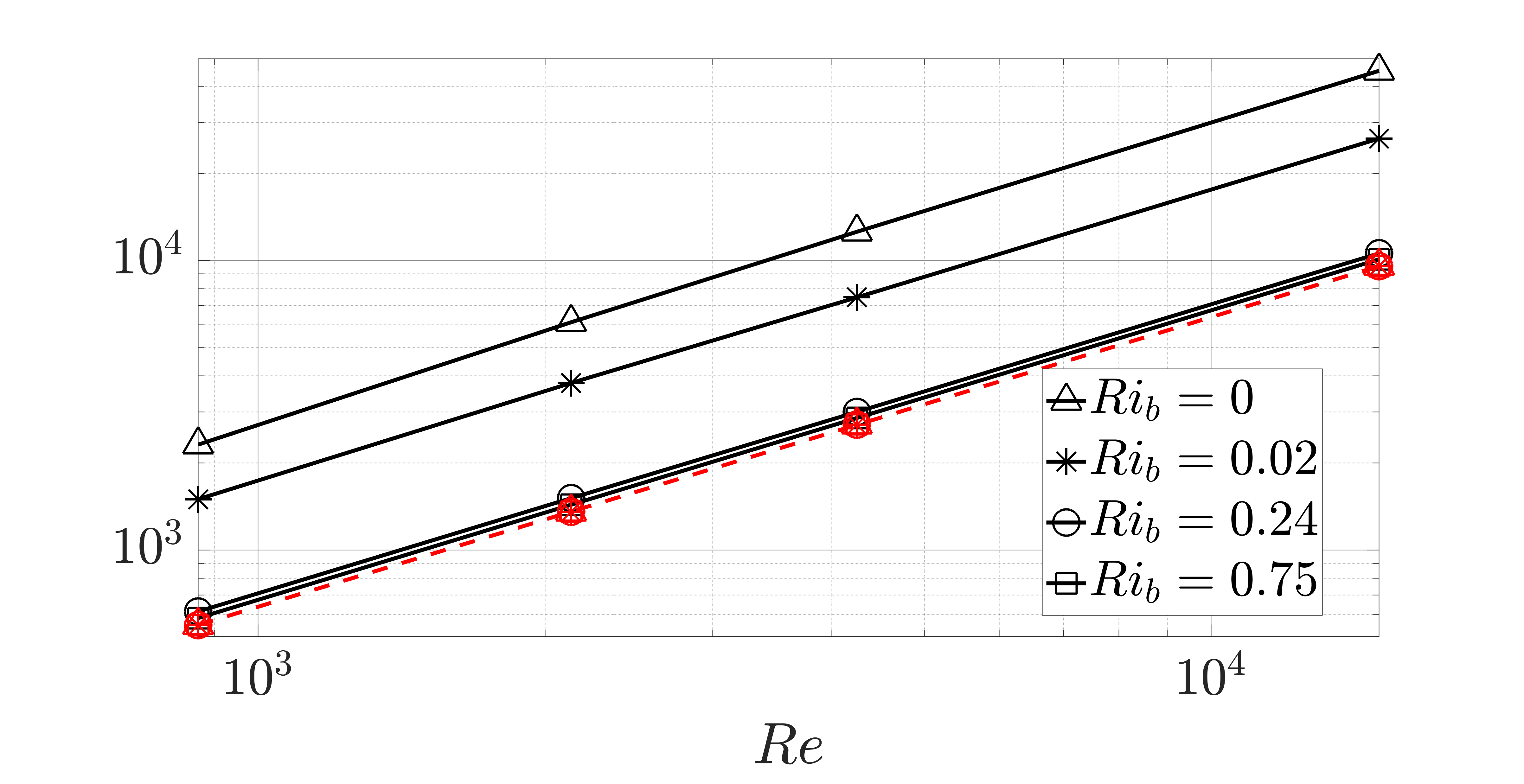}
         
    \caption{\setstretch{1.15} The dependence on $Ri_b$ and $Re$ of $\|\mathcal{H}^S_{\nabla}\|_{\mu}^M$ ($\mline\mline$, black) and $\|\mathcal{H}^{S}_{\nabla}\|_{\mu}^{sc}$ ({\color{red}$\dashed$}, red) at $Pr=0.7$. Each black marker on the lines of $\|\mathcal{H}^{S}_{\nabla}\|_{\mu}^{M}$ indicates the associated value of (a) $Re$ and (b) $Ri_b$ defined in the legend. Each red marker on the lines of $\|\mathcal{H}^{S}_{\nabla}\|_{\mu}^{sc}$ indicates the same parameter value for (a) $Re$ and (b) $Ri_b$ as the corresponding black one.}
    \label{fig:stratified_Ri_Re_variation_Pr07}
\end{figure}

The Miles-Howard theorem \citep{miles1961stability,howard1961note} implies that the laminar base flow would be linearly stable in the limits where $\nu$ and $\kappa$ both are zero if $Ri_b > 1/4$. Although the theorem is not applicable to unsteady flows with finite $Re$ and $Pr$, it  has also been observed that a `marginal' or `critical' Richardson number near this threshold value appears to emerge naturally in simulations \citep{salehipour2018self} and field measurement \citep{smyth2019self}. In the previous section, we noted that increasing $Ri_b=0.02$ to $Ri_b=0.2541$, for a fixed Reynolds number of $Re=865$, reduces the overall response and changes the types of flow structures, $(k_x, k_z)$ wavenumber pairs that exhibit the largest response, see figures \ref{fig:stratified_mu_cou_stratified}(a) and  \ref{fig:stratified_mu_high_Re_Ri_b}(b).  In this subsection, we further investigate whether this apparently marginal threshold $Ri_b \simeq 1/4$ is associated with a change in flow structures and whether this behavior is independent of $Re$. As in the previous subsection, the Prandtl number is fixed at $Pr=0.7$.

Here, we aggregate results varying over a range of $(k_x,k_z)$ wavenumber pairs in terms of the maximum value:
\begin{align}
    \|\mathcal{H}^S_{\nabla}\|_{\mu}^M:=&\underset{k_z,\,k_x}{\text{max}}\|\mathcal{H}^S_{\nabla}\|_{\mu}(k_x,k_z),
    \label{eq:stratified_H_nabla_mu_M}
\end{align}
over the wavenumber domain $k_x\in[10^{-6},10^{0.48}]$ and $k_z\in[10^{-6}, 10^{0.48}]$. Lowering the minimum value of the wave number ranges versus those considered in the previous subsection  is motivated by the observation in figure \ref{fig:stratified_mu_high_Re_Ri_b} that both the $k_x$ and $k_z$ values corresponding to the peak region decrease with increasing Reynolds and Richardson numbers. To separate streamwise-varying and streamwise-independent flow structures, we similarly evaluate
\begin{align}
        \|\mathcal{H}^S_{\nabla }\|_{\mu}^{sc}:=&\underset{k_z,\,k_x=10^{-6}}{\text{max}}\|\mathcal{H}^S_{\nabla }\|_{\mu}(k_x,k_z).
\end{align}
This quantity restricts the streamwise wavenumber to $k_x=10^{-6}$ to approximate the streamwise constant modes and computes the maximum value over $k_z\in[10^{-6}, 10^{0.48}]$, where we use an upper bound of $10^{0.48}$ rather than the larger value of $10^{1.2}$ to save computation time. This change in the upper bound was not found to affect the results since the $k_z$ associated with the maximum value was consistently found to be below this upper bound.  The restriction in streamwise wavelengths to $k_x=10^{-6}$ naturally includes the quasi-horizontal flow structures prevalent in the high-$Re$ high-$Ri_b$ regime ($k_x\approx 0$, $k_z\approx 0$) as an extreme case, but excludes streamwise varying flow structures such as the oblique turbulent bands observed in the low-$Re$ low-$Ri_b$ regime discussed in \S\ \ref{subsec:stratified_Ri_b_Re}.

Figure \ref{fig:stratified_Ri_Re_variation_Pr07} shows the variation of $\|\mathcal{H}_{\nabla}^S\|_{\mu}^M$ (solid lines) and $\|\mathcal{H}_{\nabla}^S\|_{\mu}^{sc}$ (dashed lines) with bulk Richardson number $Ri_b\in[0,6]$ and Reynolds number $Re\in[865,15000]$. The quantities $\|\mathcal{H}_{\nabla}^S\|_{\mu}^{M}$ including streamwise-varying flow structures are very close to  $\|\mathcal{H}_{\nabla}^S\|_{\mu}^{sc}$ when $Ri_b \gtrsim 1/4$ for the full range of Reynolds numbers $Re\in[865,15000]$ in figure \ref{fig:stratified_Ri_Re_variation_Pr07}(a). This phenomenon is also reflected in figure \ref{fig:stratified_Ri_Re_variation_Pr07}(b), where for flows with $Ri_b=0.24$ and $Ri_b=0.75$, the curves for 
$\|\mathcal{H}_{\nabla}^S\|_{\mu}^{M}$ and  $\|\mathcal{H}_{\nabla}^S\|_{\mu}^{sc}$
largely overlap. These trends suggest that the inviscid marginal stability value $Ri_b=1/4$ predicted by the Miles-Howard theorem \citep{miles1961stability,howard1961note} for the laminar flow is apparently associated with a change in the most amplified flow structure in stratified PCF at finite $Re$ and $Pr \sim 1$.

The plots in figure \ref{fig:stratified_Ri_Re_variation_Pr07} show that the largest amplification of the streamwise-invariant modes represented by $\|\mathcal{H}_{\nabla}^S\|_{\mu}^{sc}$ (dashed lines) do not appear to be influenced by $Ri_b$ as shown by the horizontal dashed lines in figure \ref{fig:stratified_Ri_Re_variation_Pr07}(a) and the overlapping dashed lines in figure \ref{fig:stratified_Ri_Re_variation_Pr07}(b). We further explore this $Ri_b$ independence for streamwise constant flow structures by considering the limit of horizontal invariance $\partial_x(\cdot)=0$ and $\partial_z(\cdot)=0$ ($k_x=0$ and $k_z=0$), which directly results in $\partial_y v=0$ due to the continuity equation. The boundary condition $v(y=\pm 1)=0$ then directly results in $v=0$. Using these assumptions, the advection terms vanish  (i.e., $U\partial_x(\cdot)=0$, and $\boldsymbol{u}{\cdot} \boldsymbol{\nabla}(\cdot )=0$) in each momentum and density equation. The terms associated with the background shear and density gradient also vanish due to zero wall-normal velocity; i.e, $vU'=v\overline{\rho}'=0$. 

These observations lead to a simplification of the momentum and density equations in \eqref{eq:stratified_NS_All} to:
 \begin{subequations}
 \label{eq:stratified_governing_kx_kz_0}
  \begin{align}
     \partial_t u=&\frac{1}{Re}\partial_{yy} u,\;\;
     \partial_y p=-Ri_b \rho, \tag{\theequation a,b}\\
    \partial_t w=&\frac{1}{Re}\partial_{yy} w,\;\;
    \partial_t \rho=\frac{1}{RePr}\partial_{yy} \rho. \tag{\theequation c,d}
 \end{align}
  \end{subequations}
Here, we can see that horizontal momentum and density field equations are all reduced to the diffusion equation, and the wall-normal momentum equation is reduced to a balance between the buoyancy force and the vertical pressure gradient; i.e. to a hydrostatic balance. This balance suggests that the only dependence on $Ri_b$ can be absorbed into the pressure by rescaling pressure and thus does not influence the quasi-horizontal flow structures. The results presented in figures \ref{fig:stratified_mu_cou_stratified}(a) and \ref{fig:stratified_mu_high_Re_Ri_b} suggest that flow structures with $k_x=10^{-4}$ lead to the same  structured response $\|\mathcal{H}_\nabla^S\|_\mu$ at a wide range of spanwise wavenumbers $k_z\leq 1$, and this value is consistent with that of quasi-horizontal flow structures $(k_x\approx 0,k_z\approx 0)$ that are independent of $Ri_b$.

This analysis provides evidence that quasi-horizontal flow structures are associated with amplification which is independent of $Ri_b$. Instead, high $Ri_b$ (i.e. strong stratification) will reduce the amplification of other horizontally-varying flow structures such as the oblique turbulent bands observed in the low-$Re$ low-$Ri_b$ intermittent regime of stably stratified PCF. Furthermore, it appears in figure \ref{fig:stratified_Ri_Re_variation_Pr07}(b) that quasi-horizontal flow structures are also increasingly amplified as $Re$ increases with scaling law $\|\mathcal{H}_{\nabla}^S\|_{\mu}^{sc}\sim Re$. This behavior indicates that quasi-horizontal flow structures may prefer a high-$Re$ high-$Ri_b$ regime. In \S\ \ref{sec:stratified_Re_Pr_scaling_passive_limit},  we further explore this  scaling law $\|\mathcal{H}_{\nabla}^S\|_{\mu}^{sc}\sim Re$ by developing analytical scaling arguments for $\|\mathcal{H}_{\nabla}^S\|_{\mu}$ in unstratified and streamwise-invariant flow.

\subsection{Effects of low and high $Pr$}
\label{subsec:stratified_Ri_b_Pr}

\begin{figure}

    \centering
    \includegraphics[width=0.49\textwidth]{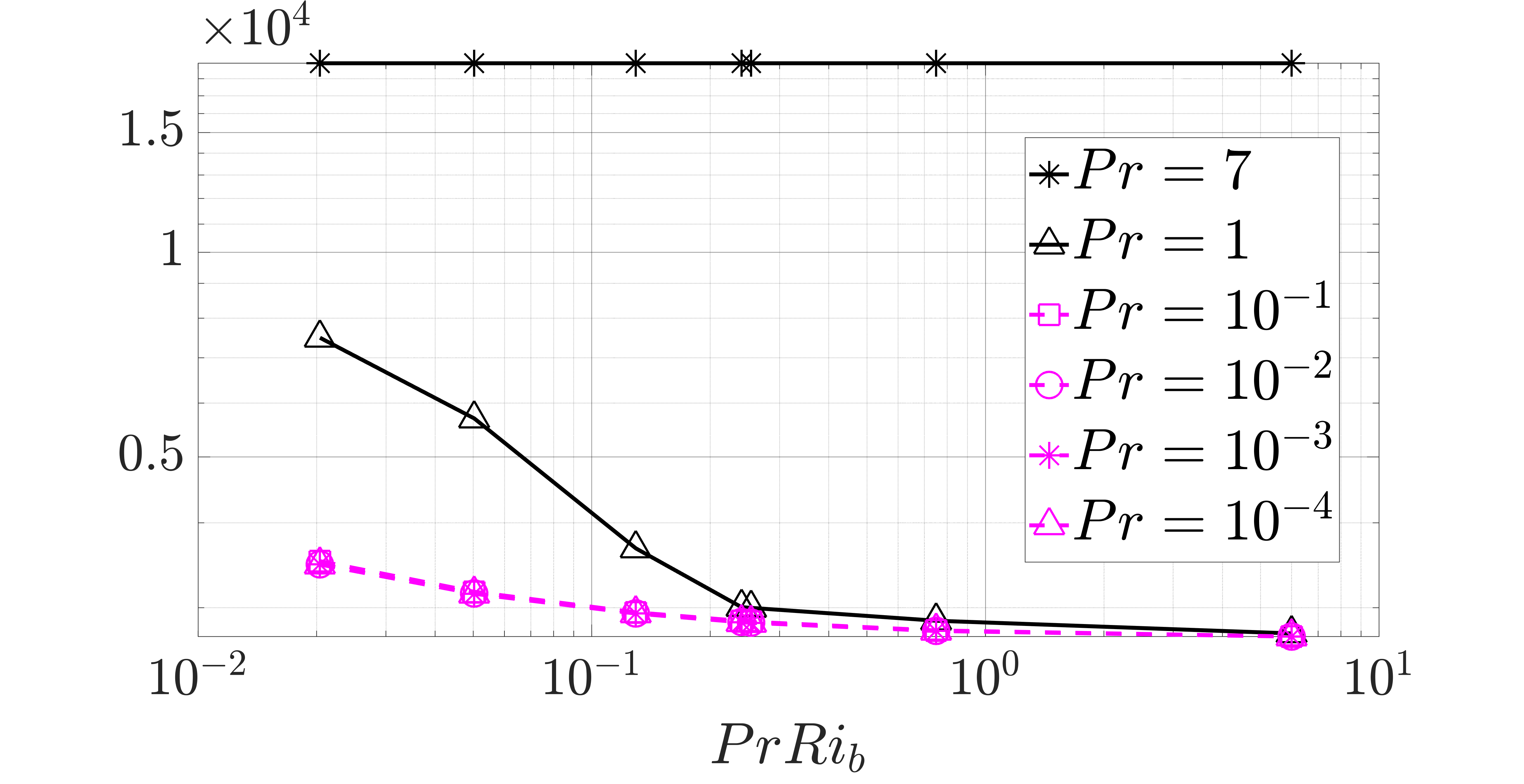}

    \caption{\setstretch{1.15} The dependence on $PrRi_b$ of $\|\mathcal{H}^S_{\nabla}\|_{\mu}^M$ at $Re=4250$.  }
    \label{fig:stratified_Ri_Pr_Ri_b_Re4250}
\end{figure}

\begin{figure}

	(a) \hspace{0.49\textwidth} (b)

    \centering
    \includegraphics[width=0.49\textwidth]{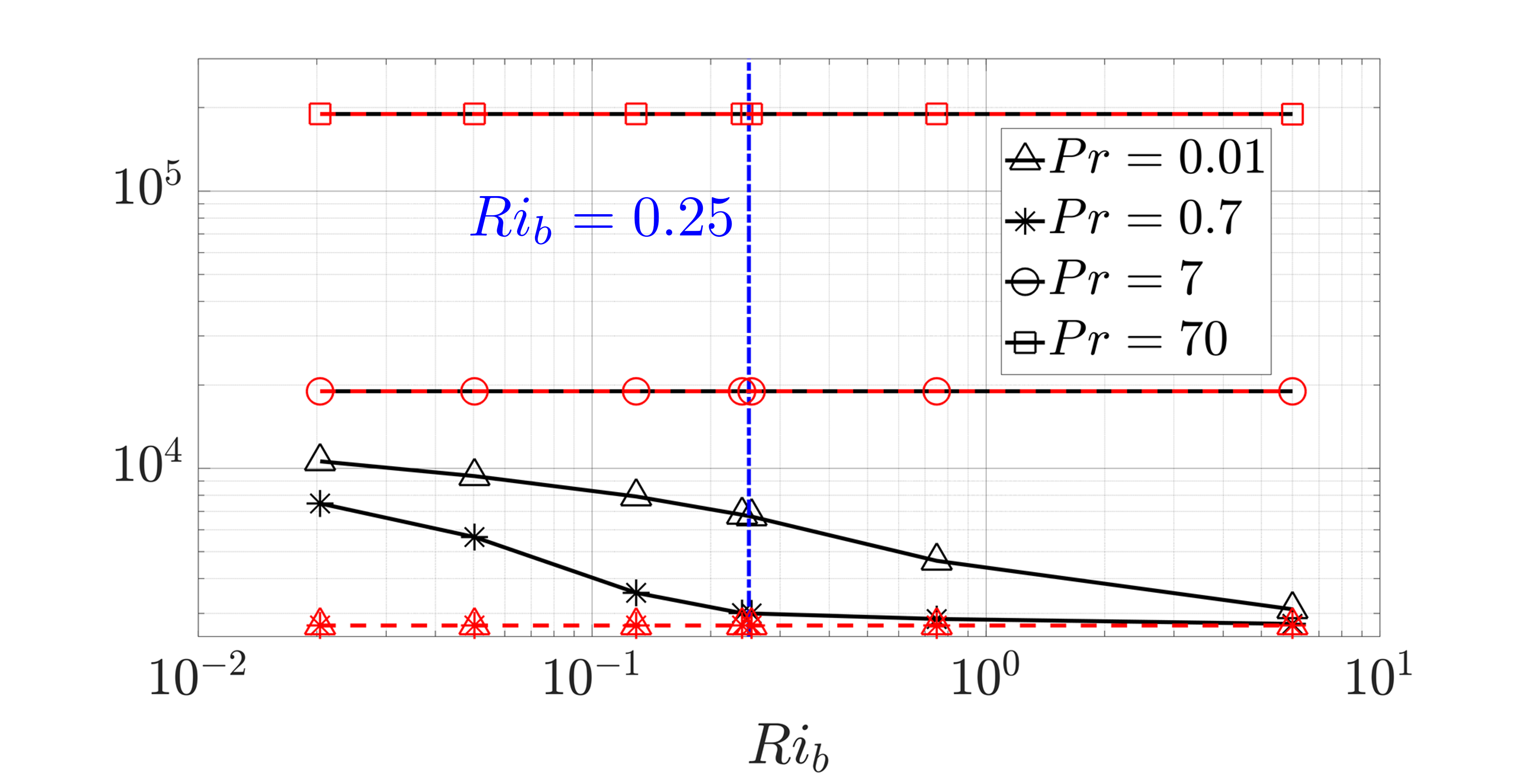}
    \includegraphics[width=0.49\textwidth]{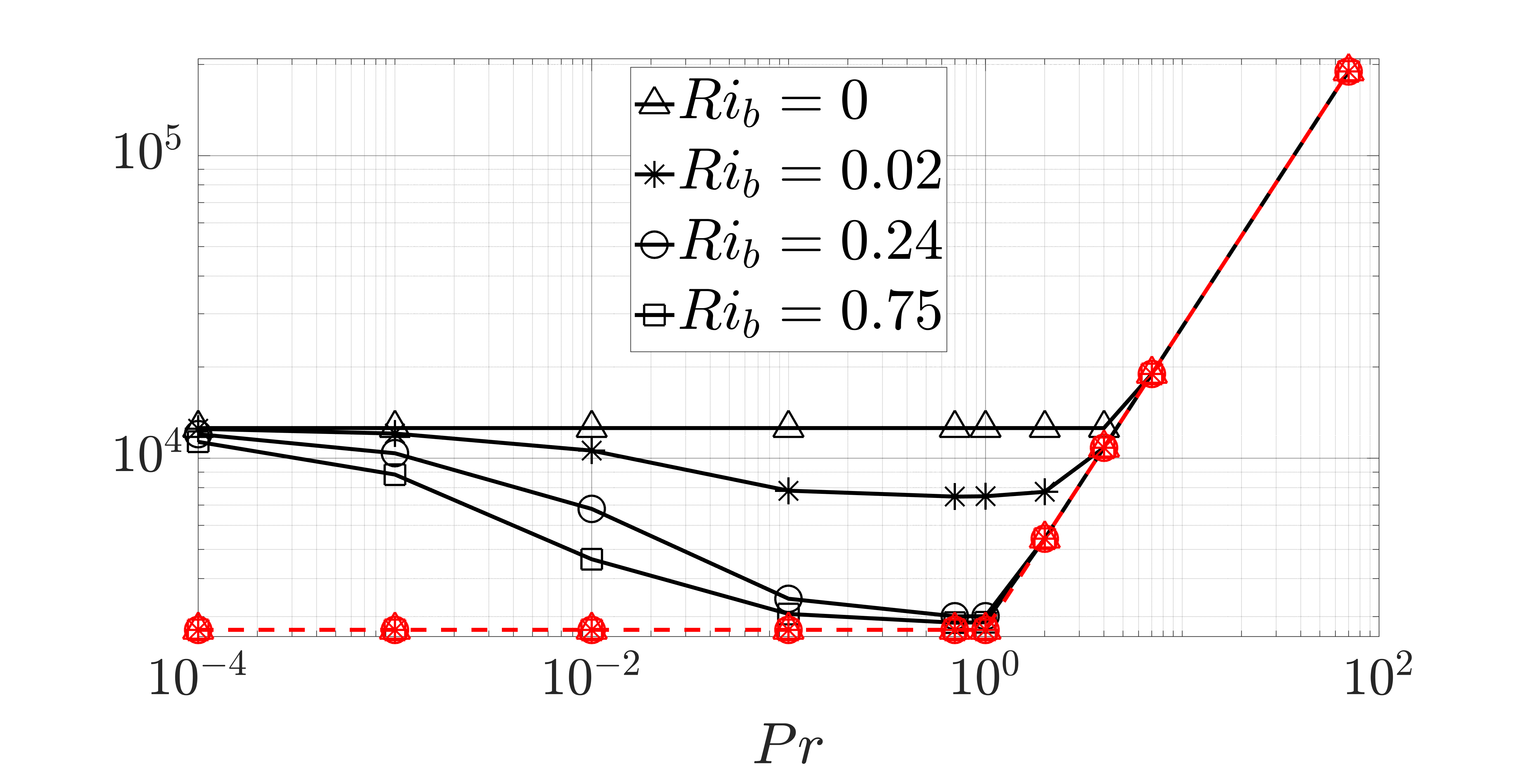}

    \caption{\setstretch{1.15} The dependence on $Ri_b$ and $Pr$ of $\|\mathcal{H}^S_{\nabla}\|_{\mu}^M$ ($\mline\mline$, black) and $\|\mathcal{H}^{S}_{\nabla}\|_{\mu}^{sc}$ ({\color{red}$\dashed$}, red) at $Re=4250$. Each black marker on the lines of $\|\mathcal{H}^S_{\nabla}\|_{\mu}^M$ indicates the associated value of (a) $Pr$ and (b) $Ri_b$ indicated in the legend. Each red marker on the lines of $\|\mathcal{H}^{S}_{\nabla}\|_{\mu}^{sc}$ indicates the same parameter as the corresponding black marker.}
    \label{fig:stratified_Ri_Pr_variation_Re4250}
\end{figure}

\begin{figure}

	\hspace{0.05\textwidth}(a) \hspace{0.28\textwidth} (b) \hspace{0.28\textwidth} (c)

    \centering
    
    \includegraphics[width=\textwidth]{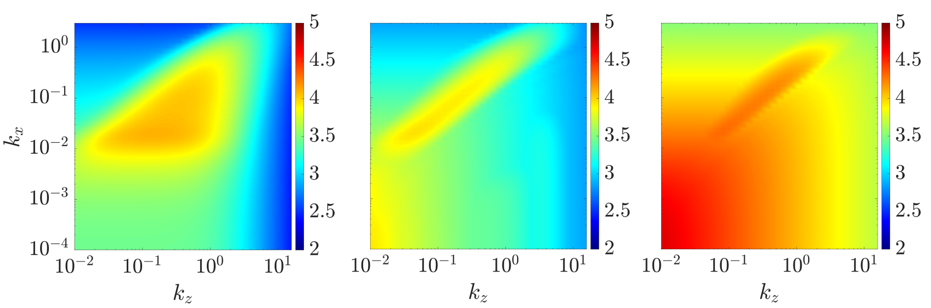}
    
    \caption{$\text{log}_{10}[\|\mathcal{H}^S_{\nabla}\|_{\mu}(k_x,k_z)]$ at $Re=4250$, $Ri_b=0.02$ with three different Prandtl numbers at (a) $Pr=10^{-4}$, (b) $Pr=7$, and (c) $Pr=70$.}
    \label{fig:stratified_Re_4250_Ri_0.02_Pr_7_70}
\end{figure} 

The Prandtl number is known to play an important role in the types of flow structures characterizing stratified PCF  \citep{zhou2017diapycnal,zhou2017self,taylor2017multi,langham2020stably}. The Prandtl number also varies over a wide range in different applications. For example, $Pr\ll 1$ is relevant for  flow in the stellar interior; see e.g., \citep{garaud2021journey}, while Prandtl number $Pr=7$ corresponds to thermally-stratified water. The Schmidt number (the analogous parameter to the Prandtl number for compositionally-induced density variations) for salt-stratified water is significantly larger $Sc \simeq 700$. Moreover, the Prandtl number is obviously not well-defined under the inviscid and nondiffusive assumptions of the Miles-Howard theorem. In this subsection, we explore the effect of low or high Prandtl numbers on flow structures. Here, we keep the Reynolds number fixed at $Re=4250$ following \citet{zhou2017self,zhou2017diapycnal}. In order to resolve fully the additional scales introduced at high $Pr$, we increase the number of wall-normal grid points to $N_y=120$ at $Pr=70$, which is chosen as a more computationally accessible `large' value, as previously considered  by  \citet{zhou2017self,zhou2017diapycnal}.

We first investigate the effect of low $Pr$. The cross-channel density profiles of exact coherent structures in stratified PCF were shown to match in flows with the same $PrRi_b$ at $Pr\in [10^{-4},10^{-2}]$ \citep[figure 3]{langham2020stably}. This combined measure $PrRi_b$ has been proposed as the natural control parameter for stably stratified shear flows at the low Prandtl number limit $Pr\ll 1$ \citep{lignieres1999small,garaud2015stability}. In order to further explore this dependence, we plot $\|\mathcal{H}_{\nabla}^S\|_{\mu}^M$ as a function of $PrRi_b$ for Prandtl numbers in the range $Pr\in [10^{-4}, 7]$ in figure \ref{fig:stratified_Ri_Pr_Ri_b_Re4250}. Here, the results $\|\mathcal{H}_{\nabla}^S\|_{\mu}^M$ for $Pr\in [10^{-4},10^{-1}]$ (magenta dashed lines) again show a natural matching dependence on $PrRi_b$. This behavior breaks down for flows with $Pr\geq 1$ as shown in figure \ref{fig:stratified_Ri_Pr_Ri_b_Re4250}. A similar end to the region of matched dependence on $PrRi_b$ alone is observed in studies using exact coherent structures, where the density profile at $Pr= 0.1$ deviates from the matching profiles for flows with $Pr\in [10^{-4}, 10^{-2}]$ yet fixed $Pr Ri_b$ \citep[figure 3]{langham2020stably}.  

In figure \ref{fig:stratified_Ri_Pr_variation_Re4250}, we plot $\|\mathcal{H}^S_{\nabla}\|_{\mu}^M$ and $\|\mathcal{H}^S_{\nabla}\|_{\mu}^{sc}$ as a function of $Ri_b$ and $Pr$ over the respective ranges $Ri_b\in[0,6]$ and  $Pr\in[10^{-4},70]$. Figure \ref{fig:stratified_Ri_Pr_variation_Re4250}(a) shows that the marginal stability value $Ri_b=1/4$ is not associated with any significant changes in the types of flow features undergoing the largest amplification for flows with $Pr=0.01$ ($\triangle$). Figure \ref{fig:stratified_Ri_Pr_variation_Re4250}(b) further suggests that flows with smaller $Pr$ require a larger $Ri_b$ to reduce the amplification of streamwise varying flow structures to the same level as streamwise constant structures. This behavior is consistent with the observation that the exact coherent structures in the low $Pr$ limit require a larger $Ri_b$ to be affected by stratification in PCF \citep{langham2020stably}. Figure \ref{fig:stratified_Ri_Pr_variation_Re4250} further shows that, for flows with high $Pr$,  the quantities $\|\mathcal{H}^S_{\nabla}\|_{\mu}^M$ and $\|\mathcal{H}^S_{\nabla}\|_{\mu}^{sc}$ are the same over a wide range of  $Ri_b\in [0,6]$. In particular,  the horizontal lines plotted in figure \ref{fig:stratified_Ri_Pr_variation_Re4250}(a) for flows with different $Ri_b$ collapse to one line  in the high $Pr\gg 1$ regime shown in figure \ref{fig:stratified_Ri_Pr_variation_Re4250}(b). This observation is also consistent with \citet{langham2020stably} who noted that in the high Prandtl number limit $Pr\gg 1$, the effect of increasing $Ri_b$ is mitigated.

In order to investigate in isolation the effect of Prandtl number on the amplification of each wavenumber pair $(k_x,k_z)$, in figure \ref{fig:stratified_Re_4250_Ri_0.02_Pr_7_70} we plot $\|\mathcal{H}^S_{\nabla}\|_{\mu}(k_x,k_z)$ for flows with (a) $Pr=10^{-4}$, (b) $Pr=7$ and (c) $Pr=70$. The peak region in figure \ref{fig:stratified_Re_4250_Ri_0.02_Pr_7_70}(a) at $Pr=10^{-4}$ resembles the shape of the peak region for unstratified PCF \citep[figure 4(a)]{liu2021structuredJournal}. We have also computed the results for unstratified PCF at the same Reynolds number $Re=4250$ (not shown here), and we find almost the same results as shown in figure \ref{fig:stratified_Re_4250_Ri_0.02_Pr_7_70}(a). This similarity suggests that for the same bulk Richardson number, a lower Prandtl number will result in a weakening of the stabilizing effect of stratification. Comparing $\|\mathcal{H}_\nabla^S\|_{\mu}$ in figure \ref{fig:stratified_mu_high_Re_Ri_b}(a) at $Pr=0.7$ with the same quantity at $Pr=7$ and $Pr=70$, respectively shown in figures \ref{fig:stratified_Re_4250_Ri_0.02_Pr_7_70}(b) and \ref{fig:stratified_Re_4250_Ri_0.02_Pr_7_70}(c), we notice that the amplification associated with the wavenumber pair $k_x=10^{-4}$ and $k_z=10^{-2}$ increases with $Pr$. More specifically, the value of $\|\mathcal{H}^S_{\nabla}\|_{\mu}$ associated with $k_x=10^{-4}$ and $k_z=10^{-2}$ becomes comparable to the values associated with the $k_x\approx 10^{-2}$ and $k_z\approx 10^{-1}$ at $Pr=7$ as shown in \ref{fig:stratified_Re_4250_Ri_0.02_Pr_7_70}(b). The wavenumber pair $k_x=10^{-4}$ and $k_z=10^{-2}$ is associated with the largest magnitude over $(k_x,k_z)$ contour region at $Pr=70$ as shown in figure \ref{fig:stratified_Re_4250_Ri_0.02_Pr_7_70}(c).

\begin{figure}
    \centering

    (a)$\|\mathcal{H}^S_{u x}\|_{\infty}$ \hspace{0.63in}
    (b)$\|\mathcal{H}^S_{v y}\|_{\infty}$ \hspace{0.57in}
    (c)$\|\mathcal{H}^S_{w z}\|_{\infty}$ \hspace{0.55in}
    (d)$\|\mathcal{H}^S_{\rho\rho}\|_{\infty}$
\includegraphics[width=\textwidth]{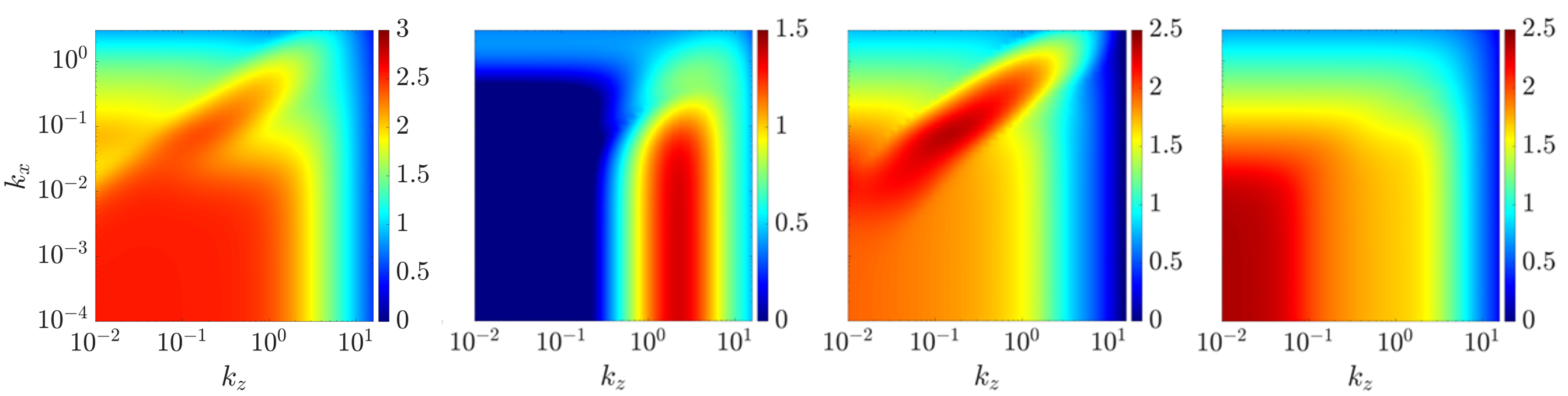}

    \caption{Componentwise values of $\text{log}_{10}[\|\mathcal{H}^S_{ij}\|_{\infty}](k_x,k_z)$ at $Re=865$, $Ri_b=0.02$, $Pr=0.7$ ($ij=ux,vy,wz,\rho\rho$).  Note the range of the colorbar for each panel is modified based on the maximum value of the particular quantity.}
    \label{fig:stratified_stratified_PCF_componentwise_H_inf_Re_865_Ri_0.02_Pr_0.7}
\end{figure}

\begin{figure}
    \centering

(a)$\|\mathcal{H}^S_{u x}\|_{\infty}$ \hspace{0.12\textwidth}
    (b)$\|\mathcal{H}^S_{v y}\|_{\infty}$ \hspace{0.11\textwidth}
    (c)$\|\mathcal{H}^S_{wz}\|_{\infty}$ \hspace{0.11\textwidth}
    (d)$\|\mathcal{H}^S_{\rho\rho}\|_{\infty}$
\includegraphics[width=\textwidth]{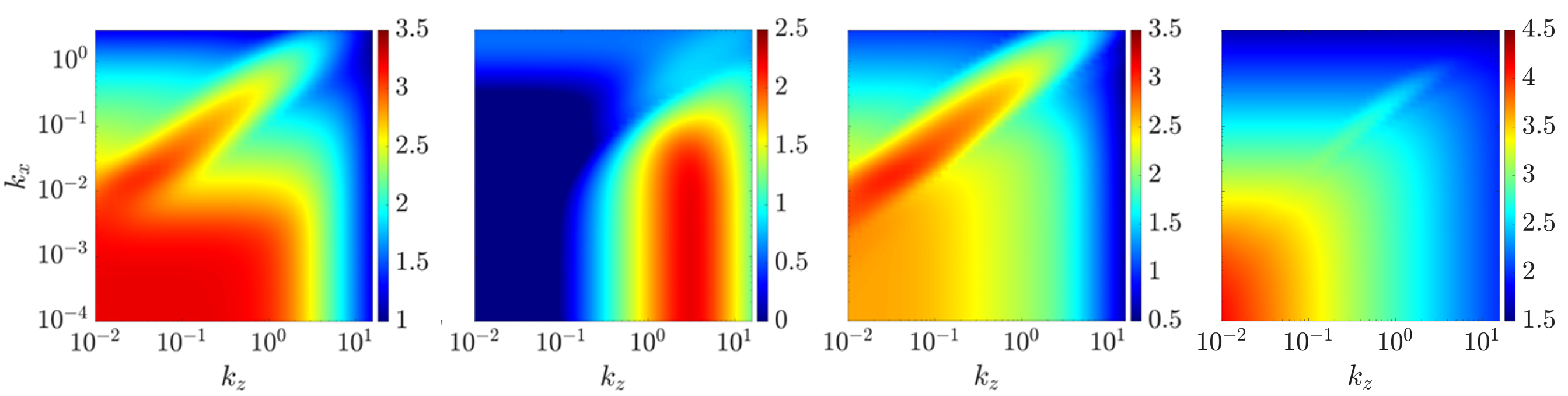}

    \caption{Componentwise values of $\text{log}_{10}[\|\mathcal{H}^S_{ij}\|_{\infty}](k_x,k_z)$ at $Re=4250$, $Ri_b=0.02$, $Pr=70$ ($ij=ux,vy,wz$, and $\rho\rho$). Note the range of the colorbar for each panel is modified based on the maximum value of the particular quantity.}
    \label{fig:stratified_stratified_PCF_componentwise_H_inf_Re_4250_Ri_0.02_Pr_70}
\end{figure}

The quasi-horizontal flow structures ($k_x\approx 0$, $k_z\approx 0$) observed in flows at high $Pr$ have different features from those previously observed in the high-$Re$, high-$Ri_b$ regime (e.g., results for flows with $Re=52630$ and $Ri_b=0.15$ shown in figure \ref{fig:stratified_mu_high_Re_Ri_b}(c)) and described in \S\ \ref{subsec:stratified_Ri_b_Re}. This indicates that a new type of quasi-horizontal flow structure appears in flows with sufficiently high $Pr$. The appearance of this flow structure at a high $Pr$ suggests that this flow structure is associated with fluctuations in the density field. This  can be further explored by isolating the input--output pathway for each component of the momentum and density equations, i.e. inputs $f_x,\, f_y, \, f_z,\, f_{\rho}$ in  \eqref{eq:stratified_f_nonlinear_all} to outputs $u$, $v$, $w$, and $\rho$. These input--output pathways can be studied through the definition of operators $\mathcal{H}^S_{ij}$, where $j$ defines the forcing input component ($j=x,y,z,\rho$) and $i=u,v,w,\rho$ describes each velocity or density output component:
\begin{equation}\mathcal{H}^S_{ij}=\widehat{\mathcal{C}}^S_{i}\left(\text{i}\omega \mathcal{I}_{3\times 3}-\widehat{\mathcal{A}}^S\right)^{-1}\widehat{\mathcal{B}}^S_{j}\label{eq:stratified_H_ij}
\end{equation}
with 
\begingroup
\allowdisplaybreaks
\begin{subequations}
\label{eq:stratified_H_componentwise}
\begin{align}
     \mathcal{\widehat{B}}^S_x:=&\mathcal{\widehat{B}}^S\begin{bmatrix}
    \mathcal{I} & 0 & 0 & 0
    \end{bmatrix}^{\text{T}},   \;\; \mathcal{\widehat{B}}^S_y:=\mathcal{\widehat{B}}^S\begin{bmatrix}
    0 & \mathcal{I} & 0 & 0
    \end{bmatrix}^{\text{T}}, \tag{\theequation a,b}\\
    \mathcal{\widehat{B}}^S_z:=&\mathcal{\widehat{B}}^S\begin{bmatrix}
    0 & 0 & \mathcal{I} & 0
    \end{bmatrix}^{\text{T}},\;\;  \mathcal{\widehat{B}}^S_\rho:=\mathcal{\widehat{B}}^S\begin{bmatrix}
    0 & 0 & 0 & \mathcal{I}     \end{bmatrix}^{\text{T}},\tag{\theequation c,d}\\ \mathcal{\widehat{C}}^S_u:=&\begin{bmatrix}\mathcal{I} & 0 & 0 & 0
    \end{bmatrix}   \mathcal{\widehat{C}}^S ,\;\;\;\;\;\mathcal{\widehat{C}}^S_v:=\begin{bmatrix}0 & \mathcal{I} & 0 & 0
    \end{bmatrix}   \mathcal{\widehat{C}}^S,\tag{\theequation e,f}\\
    \;\;\;\;\;\mathcal{\widehat{C}}^S_w:=&\begin{bmatrix}0 & 0 & \mathcal{I} & 0
    \end{bmatrix}   \mathcal{\widehat{C}}^S,
    \;\;\;\;\;\mathcal{\widehat{C}}^S_\rho:=\begin{bmatrix}0 & 0 & 0 & \mathcal{I}
    \end{bmatrix}   \mathcal{\widehat{C}}^S. \tag{\theequation g,h}
\end{align}
\end{subequations}
\endgroup

Figures \ref{fig:stratified_stratified_PCF_componentwise_H_inf_Re_865_Ri_0.02_Pr_0.7} and \ref{fig:stratified_stratified_PCF_componentwise_H_inf_Re_4250_Ri_0.02_Pr_70} present quantities $\|\mathcal{H}^S_{ij}\|_{\infty}$ $ij=ux,vy,wz$, and $\rho \rho$ for the control parameters ($Re=865, Ri_b=0.02, Pr=0.7)$, and $(Re=4250,Ri_b=0.02, Pr=70)$ respectively.  The combined effect of these four panels associated with the input and output in the same component resembles the shape of $\|\mathcal{H}_{\nabla}^S\|_{\mu}$ at the same flow regime in figures \ref{fig:stratified_mu_cou_stratified}(a) and \ref{fig:stratified_Re_4250_Ri_0.02_Pr_7_70}(c). This correspondence is because the structured feedback interconnections in \eqref{eq:stratified_f_uncertain_model}-\eqref{eq:stratified_f_density_uncertain_model} constrain the permissible feedback pathway. In figure \ref{fig:stratified_stratified_PCF_componentwise_H_inf_Re_4250_Ri_0.02_Pr_70}, panel (d) $\|\mathcal{H}_{\rho \rho}^S\|_{\infty}$ (associated with input $f_\rho$ and output $\rho$) is significantly larger than other panels for a flow with $Pr=70$ suggesting the strong role of density in the amplification for this parameter range. We can further isolate each component of the frequency response operator $\mathcal{H}_{\nabla}^S$ by defining:
\begin{align}
    \mathcal{H}_{\nabla ij}^S=\widehat{\boldsymbol{\nabla}}\mathcal{H}_{ij}^S.
    \label{eq:stratified_H_nabla_ij}
\end{align}
The values $\|\mathcal{H}_{\nabla ij}^S\|_{\infty}$, not shown here for brevity, show qualitatively similar behavior to $\|\mathcal{H}^S_{ij}\|_{\infty}$  as plotted in figures \ref{fig:stratified_stratified_PCF_componentwise_H_inf_Re_865_Ri_0.02_Pr_0.7} and \ref{fig:stratified_stratified_PCF_componentwise_H_inf_Re_4250_Ri_0.02_Pr_70}. This componentwise analysis demonstrates that the quasi-horizontal flow structures appearing at high $Pr$ are associated with density fluctuations. The appearance of this type of quasi-horizontal flow structure associated with density fluctuations in a high $Pr$ regime   is  qualitatively consistent with  previous observations that  sharp density gradients or even density `staircases' 
can be observed when $Pr$ is increased \citep{zhou2017diapycnal,taylor2017multi}.

\section{Scaling for density-associated flow structures when $Pr\gg 1$}
\label{sec:stratified_Re_Pr_scaling_passive_limit}

The previous subsection reveals the appearance of quasi-horizontal flow structures associated with density fluctuations in the $Pr\gg 1$ limit. In this section, we construct an analytical scaling of $\|\mathcal{H}_{\nabla}^S\|_{\mu}$ in terms of $Re$ and $Pr$ to provide further evidence that such flow structures prefer the $Pr\gg 1$ regime. The analytical scaling in terms of $Re$ and $Pr$ can also further provide insight into high $Re$ and $Pr$ flow regimes beyond the current computation range achievable through direct numerical simulations.

We assume streamwise-invariance ($k_x=0$) and unstratified flow ($Ri_b=0$) to facilitate analytical derivation. The importance of streamwise-invariant flow structures is suggested by the quasi-horizontal flow structures ($k_x\approx 0$ and $k_z\approx 0$), which are nearly streamwise constant. The independence with respect to variations in $Ri_b$ of the amplification of streamwise-invariant flow structures $\|\mathcal{H}_{\nabla}^S\|^{sc}_{\mu}$ shown in figures \ref{fig:stratified_Ri_Re_variation_Pr07} and \ref{fig:stratified_Ri_Pr_variation_Re4250} and the analysis of  \citet{langham2020stably} suggest that $Ri_b=0$ (i.e., density fluctuations can be treated as a passive scalar) is a reasonable regime to consider to obtain further insight. The analytically derived $Re$ and $Pr$ scalings of each component of $\mathcal{H}_{ij}^S$ in \eqref{eq:stratified_H_ij} and $\mathcal{H}_{\nabla ij}^S$ in \eqref{eq:stratified_H_nabla_ij} are presented in theorem \ref{thm:stratified_scaling_Re_Pr}(a) and (b), respectively. 

\begingroup
\allowdisplaybreaks
\begin{thm}
\label{thm:stratified_scaling_Re_Pr}
Consider streamwise-invariant ($k_x=0$) unstratified ($Ri_b=0$) plane Couette flow with a passive scalar `density' field.

(a) Each component of $\|\mathcal{H}^S_{ij}\|_{\infty}$ ($i=u,v,w,\rho$ and $j=x,y,z,\rho$) scales as:
\begin{align}
    & \begin{bmatrix}
    \|\mathcal{H}^S_{ux}\|_{\infty} & \|\mathcal{H}^S_{uy}\|_{\infty} & \|\mathcal{H}^S_{uz}\|_{\infty} & \|\mathcal{H}^S_{u\rho}\|_{\infty}\\
    \|\mathcal{H}^S_{vx}\|_{\infty} & \|\mathcal{H}^S_{vy}\|_{\infty} & \|\mathcal{H}^S_{vz}\|_{\infty} & \|\mathcal{H}^S_{v\rho}\|_{\infty}\\
    \|\mathcal{H}^S_{wx}\|_{\infty} & \|\mathcal{H}^S_{wy}\|_{\infty} & \|\mathcal{H}^S_{wz}\|_{\infty} & \|\mathcal{H}^S_{w\rho}\|_{\infty}\\
    \|\mathcal{H}^S_{\rho x}\|_{\infty} & \|\mathcal{H}^S_{\rho y}\|_{\infty} & \|\mathcal{H}^S_{\rho z}\|_{\infty} & \|\mathcal{H}^S_{\rho \rho}\|_{\infty}\\
\end{bmatrix} \nonumber \\
=&\begin{bmatrix}
    Re\, h^S_{ux}(k_z) & Re^2\, h^S_{uy}(k_z) & Re^2\, h^S_{uz}(k_z) & 0\\
    0 & Re\,h^S_{vy}(k_z) & Re\,h^S_{vz}( k_z) & 0\\
    0 & Re\, h^S_{wy}(k_z) & Re\, h^S_{wz}(k_z) & 0\\
    0 & Re^2Pr\,h^S_{\rho y}(k_z) & Re^2Pr\,h^S_{\rho z}(k_z) & RePr\, h^S_{\rho \rho}(k_z)
    \end{bmatrix},
    \label{eq:stratified_scaling_Re_Pr_a}
\end{align}
where functions $h^S_{ij}(k_z)$ are independent of $Re$ and $Pr$.

(b) Each component of $\|\mathcal{H}^S_{\nabla ij}\|_{\infty}$ ($i=u,v,w,\rho$ and $j=x,y,z,\rho$) scales as:
\begin{align}
    & \begin{bmatrix}
    \|\mathcal{H}^S_{\nabla ux}\|_{\infty} & \|\mathcal{H}^S_{\nabla uy}\|_{\infty} & \|\mathcal{H}^S_{\nabla uz}\|_{\infty} & \|\mathcal{H}^S_{\nabla u\rho}\|_{\infty}\\
    \|\mathcal{H}^S_{ \nabla vx}\|_{\infty} & \|\mathcal{H}^S_{\nabla vy}\|_{\infty} & \|\mathcal{H}^S_{\nabla vz}\|_{\infty} & \|\mathcal{H}^S_{\nabla v\rho}\|_{\infty}\\
    \|\mathcal{H}^S_{\nabla wx}\|_{\infty} & \|\mathcal{H}^S_{\nabla wy}\|_{\infty} & \|\mathcal{H}^S_{\nabla wz}\|_{\infty} & \|\mathcal{H}^S_{\nabla w\rho}\|_{\infty}\\
    \|\mathcal{H}^S_{\nabla \rho x}\|_{\infty} & \|\mathcal{H}^S_{\nabla\rho y}\|_{\infty} & \|\mathcal{H}^S_{\nabla\rho z}\|_{\infty} & \|\mathcal{H}^S_{\nabla\rho \rho}\|_{\infty}\\
\end{bmatrix} \nonumber \\
=&\begin{bmatrix}
    Re\, h^S_{\nabla ux}(k_z) & Re^2\, h^S_{\nabla uy}(k_z) & Re^2\, h^S_{ \nabla uz}(k_z) & 0\\
    0 & Re\,h^S_{\nabla vy}(k_z) & Re\,h^S_{ \nabla vz}( k_z) & 0\\
    0 & Re\, h^S_{\nabla wy}(k_z) & Re \,h^S_{ \nabla wz}(k_z) & 0\\
    0 & Re^2Pr\,h^S_{\nabla\rho y}(k_z) & Re^2Pr\,h^S_{\nabla\rho z}(k_z) & RePr\, h^S_{\nabla\rho \rho}(k_z)
    \end{bmatrix},
    \label{eq:stratified_scaling_Re_Pr_b}
\end{align}
where functions $h^S_{\nabla ij}(k_z)$ are independent of $Re$ and $Pr$. 
\end{thm}

\endgroup

The first three columns and three rows presented in equation \eqref{eq:stratified_scaling_Re_Pr_a} are the same as those derived in \citet[theorem 11]{jovanovic2004modeling} for unstratified wall-bounded shear flows with no passive scalar field. The details of the proof are presented in Appendix \ref{appendix:stratified_proof_scaling_Re_Pr}. These results demonstrate that the $Pr$ only contributes to the scaling associated with the density field (here of course assumed to be a passive scalar); i.e. the bottom rows of equations \eqref{eq:stratified_scaling_Re_Pr_a} and \eqref{eq:stratified_scaling_Re_Pr_b} corresponding to the density output. We also note that the rightmost columns of equations \eqref{eq:stratified_scaling_Re_Pr_a} and \eqref{eq:stratified_scaling_Re_Pr_b} show that the forcing in the density equation $f_{\rho}$ does not influence the output corresponding to velocity components $u$, $v$, and $w$, which is consistent with the assumption that $Ri_b=0$, in that the density perturbation behaves as a passive scalar.

The effect of imposing a componentwise structure of nonlinearity within the feedback is analogous to the effect seen in unstratified PCF \citep[section 3.3]{liu2021structuredJournal}. The imposed correlation between each component of the modeled forcing $f_{x,\xi}$, $f_{y,\xi}$, $f_{z,\xi}$, $f_{\rho,\xi}$, and the respective velocity and density components $u$, $v$, $w$, $\rho$ constrain the influence of the forcing to its associated component of the velocity or density field. Thus, the overall scaling of $\|\mathcal{H}_{\nabla}^S\|_{\mu}$ is related to the worst-case scaling of the diagonal terms in equation \eqref{eq:stratified_scaling_Re_Pr_b} in theorem \ref{thm:stratified_scaling_Re_Pr}. The concept is formalized in theorem \ref{lemma:stratified_mu_componentwise_inf}, and we relegate the details of the proof to Appendix \ref{appendix:stratified_proof_scaling_Re_Pr}. 
\begin{thm}
Given a wavenumber pair $(k_x,k_z)$. 
\begin{align}
    \|\mathcal{H}_{\nabla}^S\|_{\mu}\geq \max[\|\mathcal{H}_{\nabla ux}^S\|_{\infty},\|\mathcal{H}_{\nabla vy}^S\|_{\infty}, \|\mathcal{H}_{\nabla wz}^S\|_{\infty}, \|\mathcal{H}_{\nabla \rho \rho}^S\|_{\infty}].
    \label{eq:stratified_mu_larger_than_all_diagonal_component}
\end{align}
\label{lemma:stratified_mu_componentwise_inf}
\end{thm}

We can combine results in theorem \ref{thm:stratified_scaling_Re_Pr}(b) and theorem \ref{lemma:stratified_mu_componentwise_inf} to obtain the scaling of $\|\mathcal{H}_{\nabla}^S\|_{\mu}$ in corollary \ref{thm:stratified_scaling_mu}:
\begin{corollary}
Consider streamwise-invariant ($k_x=0$) unstratified ($Ri_b=0$) plane Couette flow with a passive scalar `density' field.
\begin{align}
    \|\mathcal{H}_{\nabla}^S\|_{\mu}(0,k_z)\geq \max[Re\, h^S_{\nabla ux }(k_z), Re\, h^S_{\nabla vy }(k_z),Re\, h^S_{\nabla wz }(k_z), RePr\, h^S_{\nabla \rho \rho}(k_z)],
    \label{eq:stratified_mu_componentwise_inequality_Re_Pr}
\end{align}
where functions $h^S_{\nabla ij}(k_z)$ with $ij=ux,vy,wz,\rho\rho$ are independent of $Re$ and $Pr$. 
\label{thm:stratified_scaling_mu}
\end{corollary}

Although corollary \ref{thm:stratified_scaling_mu} provides a lower bound on  $\|\mathcal{H}_{\nabla}^S\|_{\mu}$, the numerical results suggest that $\|\mathcal{H}_{\nabla}^S\|_{\mu}$ follows the same $Re$ and $Pr$ scaling as the right-hand side of \eqref{eq:stratified_mu_componentwise_inequality_Re_Pr} in corollary \ref{thm:stratified_scaling_mu}. For example, corollary \ref{thm:stratified_scaling_mu} suggests that the lower bound of $\|\mathcal{H}_{\nabla}^S\|_{\mu}(0,k_z)$ will scale as $\sim Re$ at a fixed $Pr$, which is consistent with the red dashed lines of figure \ref{fig:stratified_Ri_Re_variation_Pr07}(b). At a fixed $Re$, corollary \ref{thm:stratified_scaling_mu} also suggests that $\|\mathcal{H}_{\nabla}^S\|_{\mu}(0,k_z)\sim Pr$ in the limit $Pr\gg 1$, but $\|\mathcal{H}_{\nabla}^S\|_{\mu}(0,k_z)$ will become independent of $Pr$ in the limit $Pr\ll 1$. This is also consistent with the numerical results shown in the red dashed lines of figure \ref{fig:stratified_Ri_Pr_variation_Re4250}(b) that suggest $\|\mathcal{H}_{\nabla}^{S}\|^{sc}_{\mu}\sim Pr$ when $Pr\gg 1$ and independently of $Pr$ when $Pr\ll 1$. For $Pr\gg 1$, theorem \ref{lemma:stratified_mu_componentwise_inf} and corollary \ref{thm:stratified_scaling_mu} further suggest that the component $\|\mathcal{H}^{S}_{\nabla \rho \rho}\|_{\infty}$ associated with the density will dominate the overall behavior of $\|\mathcal{H}_{\nabla}^S\|_{\mu}$, which is consistent with the large amplification of quasi-horizontal flow structures associated with density fluctuations $\|\mathcal{H}^S_{\rho \rho}\|_\infty$ shown in figure \ref{fig:stratified_stratified_PCF_componentwise_H_inf_Re_4250_Ri_0.02_Pr_70}(p). Corollary \ref{thm:stratified_scaling_mu} further supports the notion that the flow structures associated with density fluctuations prefer the flow regime with $Pr\gg 1$ under the assumptions of streamwise-invariant ($k_x=0$) and unstratified ($Ri_b=0$) flow.

\section{Conclusions and future work}
\label{sec:stratified_stratified_conclusion}

In this paper, we have extended the structured input--output analysis (SIOA) originally developed for unstratified wall-bounded shear flows \citep{liu2021structuredJournal} to stratified plane Couette flow (PCF). We first apply SIOA to characterize highly amplified flow structures in the intermittent regimes of stratified PCF. We first examine how variations in $Re$ and $Ri_b$ affect flow structures with $Pr=0.7$. SIOA predicts the characteristic wavelengths and angle of the oblique turbulent bands observed in very large channel size DNS of the low-$Re$ low-$Ri_b$ intermittent regime of stratified PCF \citep{deusebio2015intermittency,taylor2016new}. In the high-$Re$ high-$Ri_b$ intermittent regime, SIOA identifies quasi-horizontal flow structures resembling turbulent-laminar layers \citep{deusebio2015intermittency}. 

Having validated the ability of the SIOA approach to predict important structures in the intermittent regime, we next investigate the behavior of the flow across a range of the important control parameters $Re$, $Ri_b$ and $Pr$. Increasing $Ri_b$ is shown to reduce the amplification of streamwise varying flow structures. The results indicate that the classical  marginally stable $Ri_b= 1/4$ for the laminar base flow appears to be associated with a change in the most amplified flow structures, an observation which is robust for a wide range of $Re$ and valid at $Pr\approx 1$.

We then examine flow behavior at different $Ri_b$ and $Pr$. For flows with $Pr\ll 1$, a larger value of $Ri_b$ is required to reduce the amplification of streamwise varying flow structures to the same level as streamwise-invariant ones compared with flows with $Pr\approx 1$. The largest amplification also occurs at the same value of $Pr Ri_b$ consistent with the observation of matching averaged density profile for flows with the same value of $PrRi_b$ in the $Pr\ll 1$ regime  \citep{langham2020stably}. For flows with $Pr\gg 1$, the SIOA identifies another quasi-horizontal flow structure that is independent of $Ri_b$. By decomposing input--output pathways into each velocity and density component, we show that these quasi-horizontal flow structures for flows with $Pr\gg 1$ are associated with density fluctuations. The importance of this density-associated flow structure for flows with $Pr\gg 1$ is further highlighted through a derived analytical scaling of amplification with respect to $Re$ and $Pr$ under the assumptions that the flow is streamwise invariant ($k_x=0$) and unstratified (i.e. $Ri_b=0$ and the density behaves as a passive scalar). The above observations using SIOA distinguish two types of quasi-horizontal flow structures, one emerging in the high-$Re$ high-$Ri_b$ regime and the other one (associated with density fluctuations) emerging in the high $Pr$ regime. 

The results here suggest the promise of this computationally tractable approach in identifying horizontal length scales of prominent flow structures in stratified wall-bounded shear flows and opens up many directions for future work. For example, this framework may be extended to other stratified wall-bounded shear flows such as stratified channel flow \citep{garcia2011turbulence}, stratified open channels \citep{flores2011analysis,brethouwer2012turbulent,donda2015collapse,he2015direct,he2016development}, and the stratified Ekman layer \citep{deusebio2014numerical}, where  intermittent regimes of flow dynamics were also observed. This framework may be also extended to configurations where the background density gradient and velocity gradient are orthogonal; e.g., spanwise stratified plane Couette flow \citep{facchini2018linear,lucas2019layer} and spanwise stratified plane Poiseuille flow \citep{le2020instability}.

\section*{Acknowledgment}
C.L. and D.F.G. gratefully acknowledge support from the US National Science Foundation (NSF) through grant number CBET 1652244. C.L. also greatly appreciates the support from the Chinese Scholarship Council.

\section*{Declaration of Interests}
The authors report no conflict of interest.

\appendix

\section{Growth rate analysis}
\label{appendix:growth_rate}

\begin{figure}
    \centering
    \includegraphics[width=0.35\textwidth]{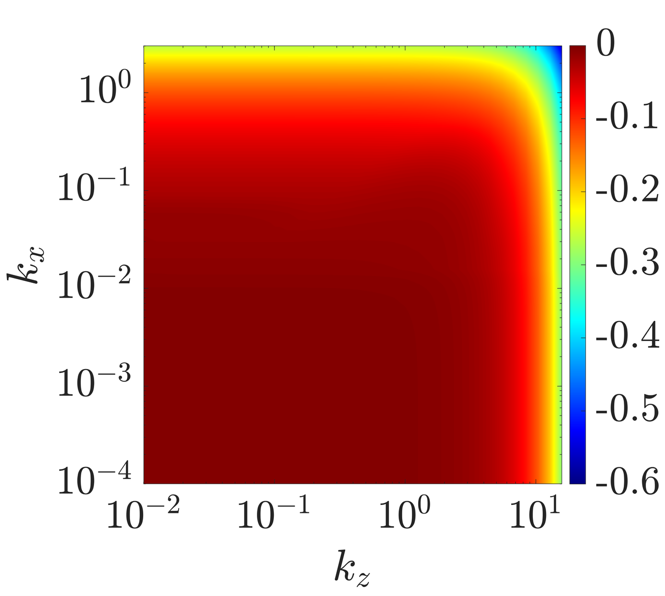}
    \caption{$R(\widehat{\mathcal{A}}^S)(k_x,k_z)$ for stratified plane Couette flow in a flow with $Re=865$, $Ri_b=0.02$, and $Pr=0.7$.}
    \label{fig:stratified_growth_rate}
\end{figure}

Here, we present the growth rate of the dynamics in equation \eqref{eq:stratified_ABC_frequency} computed as:
\begin{align}
    R[\widehat{\mathcal{A}}^S(k_x,k_z)]:=\text{max}\left\{\mathbb{R}e\left[\text{eig}\left(\widehat{\mathsfbi{A}}^S(k_x,k_z)\right)\right]\right\},
    \label{eq:stratified_growth_rate}
\end{align}
where $\text{eig}(\cdot)$ is the eigenvalue of the argument, $\mathbb{R}e[\cdot]$ represents the real part, $\text{max}\{\cdot\}$ is the maximum value of the argument, and $\widehat{\mathsfbi{A}}^S$ is the discretization of operator $\widehat{\mathcal{A}}^S$. Figure \ref{fig:stratified_growth_rate} shows the growth rate $R(\widehat{\mathcal{A}}^S)(k_x,k_z)$ in \eqref{eq:stratified_growth_rate}. Here, we observe that this modal growth rate analysis $R(\widehat{\mathcal{A}}^S)(k_x,k_z)$ cannot distinguish a preferential structure size over a wide range of wavenumbers $k_x\lesssim  1$ and $k_z\lesssim  10$, and there is no identified instability consistent with \citet{davey1977stability}. 

\section{Proof of theorems \ref{thm:stratified_scaling_Re_Pr}-\ref{lemma:stratified_mu_componentwise_inf}}

\label{appendix:stratified_proof_scaling_Re_Pr}

\subsection{Proof of theorem \ref{thm:stratified_scaling_Re_Pr}}
\begin{myproof}
The proof of theorem \ref{thm:stratified_scaling_Re_Pr} naturally follows the procedure in unstratified flow \citep{jovanovic2004modeling,Jovanovic2005,jovanovic2020bypass} and is outlined as a block diagram in figure \ref{fig:stratified_block_diagram_Re_Pr_scaling}. Under the assumption of streamwise-invariance ($k_x=0$) and taking the passive scalar limit ($Ri_b=0$) for stratified plane Couette flow (PCF) in theorem \ref{thm:stratified_scaling_Re_Pr}, the operators $\widehat{\mathcal{A}}^S$, $\widehat{\mathcal{B}}^S$, and $\widehat{\mathcal{C}}^S$ can be simplified and their non-zero elements can be defined as:

\begin{figure}
     \centering
     \includegraphics[width=\textwidth]{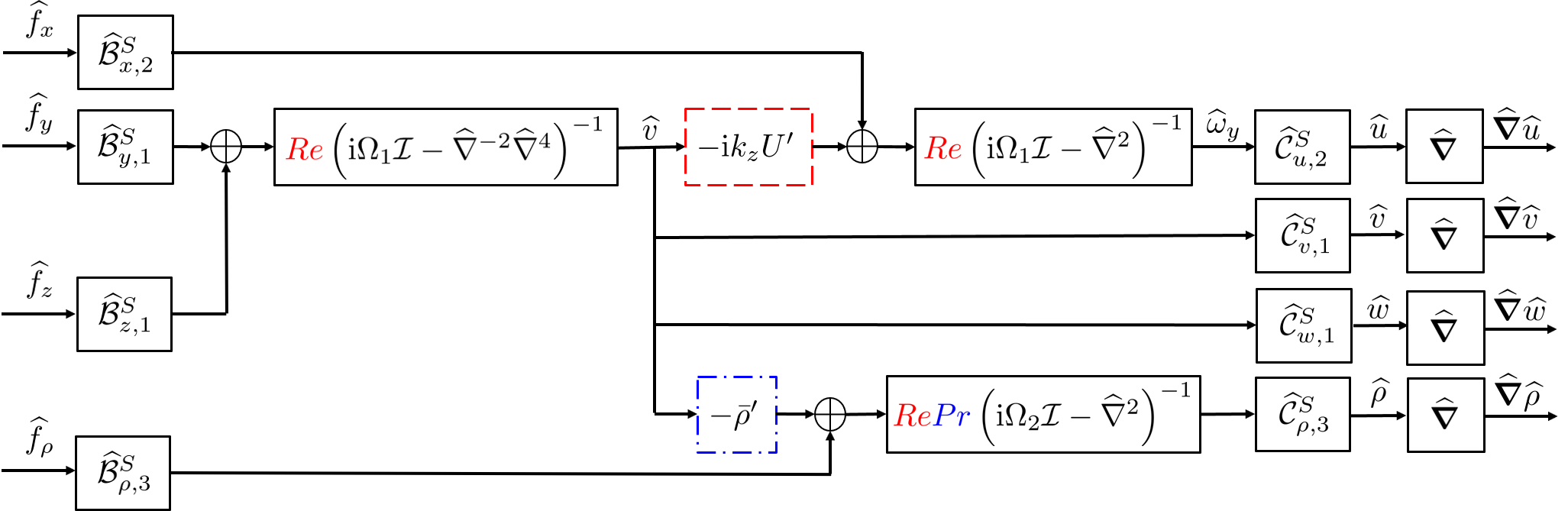}
     \caption{\setstretch{1}Block diagram of the frequency response operator that maps forcing in each momentum and density  equation to each velocity and density gradient in streamwise-invariant ($k_x=0$) PCF with density assumed to be a passive scalar (i.e. $Ri_b=0$). Here, $\Omega_1=\omega Re$ and $\Omega_2=\omega Re Pr$. The block outlined by ({\color{red}$\dashed$}, red) contributes to the scaling associated with $\|\mathcal{H}^S_{uy}\|_{\infty}$, $\|\mathcal{H}^S_{uz}\|_{\infty}$, while the block outlined by ({\color{blue}$\dashdot$}, blue) contributes to the scaling associated with $\|\mathcal{H}^S_{\rho y}\|_{\infty}$, $\|\mathcal{H}^S_{\rho z }\|_{\infty}$. }
     \label{fig:stratified_block_diagram_Re_Pr_scaling}
 \end{figure}

\begin{subequations}
\label{eq:stratified_operator_ABC_appendix_proof}
\begin{align}
    \widehat{\mathcal{A}}^S(k_x,k_z)=&\begin{bmatrix}
   \frac{\widehat{{\nabla}}^{-2}\widehat{{\nabla}}^4}{Re} & 0 & 0\\
    -\text{i}k_zU' & \frac{\widehat{{\nabla}}^2}{Re} & 0\\
    -\overline{\rho}' & 0 & \frac{\widehat{\nabla}^2}{Re Pr}
    \end{bmatrix},\\
    \mathcal{\widehat{B}}^S(k_x,k_z)=&
    \begin{bmatrix}
    0 & -k_z^2\widehat{{\nabla}}^{-2} & -\text{i}k_z\widehat{{\nabla}}^{-2} \partial_y & 0\\
    \text{i}k_z & 0 & 0 & 0\\
    0 & 0 & 0 & \mathcal{I}
    \end{bmatrix}=:\begin{bmatrix}
    0 & \widehat{\mathcal{B}}^S_{y,1} & \widehat{\mathcal{B}}^S_{z,1} & 0\\
    \mathcal{B}^S_{x,2} & 0 & 0 & 0\\
    0 & 0 & 0& \widehat{\mathcal{B}}^S_{\rho,3}
    \end{bmatrix}\\
    \mathcal{\widehat{C}}^S(k_x,k_z)=&\begin{bmatrix}
    0 & -\text{i}/k_z & 0 \\
    \mathcal{I} & 0 & 0\\
    \text{i} \partial_y/k_z & 0 & 0 \\
    0 & 0 & \mathcal{I}
    \end{bmatrix}=:\begin{bmatrix}
    0 &  \mathcal{\widehat{C}}^S_{u,2} & 0 \\
     \mathcal{\widehat{C}}^S_{v,1} & 0 & 0\\
     \mathcal{\widehat{C}}^S_{w,1} & 0 & 0 \\
    0 & 0 &  \mathcal{\widehat{C}}^S_{\rho,3} 
    \end{bmatrix}
\end{align}
\end{subequations}
We employ a matrix inverse formula for the lower triangle block matrix:
\begin{align}
        \begin{bmatrix}
    L_{11} & 0 & 0\\
    L_{21} & L_{22} & 0\\
    L_{31} & 0 & L_{33}
    \end{bmatrix}^{-1}=\begin{bmatrix}
    L_{11}^{-1} & 0 & 0\\
    -L_{22}^{-1}L_{21}L_{11}^{-1} & L_{22}^{-1} & 0\\
    -L_{33}^{-1}L_{31}L_{11}^{-1} & 0 & L_{33}^{-1}
    \end{bmatrix}
\end{align}
to compute $\left(\text{i}\omega\mathcal{I}_{3\times 3}-\widehat{\mathcal{A}}^S\right)^{-1}$. Then, we employ a change of variable $\Omega_1=\omega Re$ and $\Omega_2=\omega RePr$ to obtain componentwise frequency response operators $\mathcal{H}^S_{ij}$ with $i=u,v,w,\rho$, and $j=x,y,z,\rho$ as:
\begingroup
\allowdisplaybreaks
\begin{subequations}
\begin{align}
    \mathcal{H}^S_{ux}=&\widehat{\mathcal{C}}^S_{u,2}Re\left(\text{i}\Omega_1 \mathcal{I}-\widehat{{\nabla}}^2\right)^{-1}\widehat{\mathcal{B}}^S_{x,2},\\
    \mathcal{H}^S_{uy}=&\widehat{\mathcal{C}}^S_{u,2}Re\left(\text{i}\Omega_1 \mathcal{I}-\widehat{{\nabla}}^2\right)^{-1}(-\text{i}k_zU')Re\left(\text{i}\Omega_1 \mathcal{I}-\widehat{{\nabla}}^{-2}\widehat{{\nabla}}^{4}\right)^{-1} \widehat{\mathcal{B}}^S_{y,1},\\
    \mathcal{H}^S_{uz}=&\widehat{\mathcal{C}}^S_{u,2}Re\left(\text{i}\Omega_1 \mathcal{I}-\widehat{{\nabla}}^2\right)^{-1}(-\text{i}k_zU')Re\left(\text{i}\Omega_1 \mathcal{I}-\widehat{{\nabla}}^{-2}\widehat{{\nabla}}^{4}\right)^{-1} \widehat{\mathcal{B}}^S_{z,1},\\
    \mathcal{H}^S_{u\rho}=&0,\\
    \mathcal{H}^S_{vx}=&0,\\
    \mathcal{H}^S_{vy}=&\widehat{\mathcal{C}}^S_{v,1}Re\left(\text{i}\Omega_1 \mathcal{I}-\widehat{{\nabla}}^{-2}\widehat{{\nabla}}^{4}\right)^{-1} \widehat{\mathcal{B}}^S_{y,1},\\
    \mathcal{H}^S_{vz}=&\widehat{\mathcal{C}}^S_{v,1}Re\left(\text{i}\Omega_1 \mathcal{I}-\widehat{{\nabla}}^{-2}\widehat{{\nabla}}^{4}\right)^{-1} \widehat{\mathcal{B}}^S_{z,1},\\
    \mathcal{H}^S_{v\rho}=&0,\\   \mathcal{H}^S_{wx}=&0,\\
    \mathcal{H}^S_{wy}=&\widehat{\mathcal{C}}^S_{w,1}Re\left(\text{i}\Omega_1 \mathcal{I}-\widehat{{\nabla}}^{-2}\widehat{{\nabla}}^{4}\right)^{-1} \widehat{\mathcal{B}}^S_{y,1},\\
    \mathcal{H}^S_{wz}=&\widehat{\mathcal{C}}^S_{w,1}Re\left(\text{i}\Omega_1 \mathcal{I}-\widehat{{\nabla}}^{-2}\widehat{{\nabla}}^{4}\right)^{-1} \widehat{\mathcal{B}}^S_{z,1},\\
    \mathcal{H}^S_{w\rho}=&0,\\
    \mathcal{H}^S_{\rho x}=&0,\\
    \mathcal{H}^S_{\rho y}=&\widehat{\mathcal{C}}^S_{\rho,3}RePr\left(\text{i}\Omega_2 \mathcal{I}-\widehat{{\nabla}}^2\right)^{-1}(-\bar{\rho}')Re\left(\text{i}\Omega_1 \mathcal{I}-\widehat{{\nabla}}^{-2}\widehat{{\nabla}}^{4}\right)^{-1} \widehat{\mathcal{B}}^S_{y,1},\\
    \mathcal{H}^S_{\rho z}=&\widehat{\mathcal{C}}^S_{\rho,3}RePr\left(\text{i}\Omega_2 \mathcal{I}-\widehat{{\nabla}}^2\right)^{-1}(-\bar{\rho}')Re\left(\text{i}\Omega_1 \mathcal{I}-\widehat{{\nabla}}^{-2}\widehat{{\nabla}}^{4}\right)^{-1} \widehat{\mathcal{B}}^S_{z,1},\\
    \mathcal{H}^S_{\rho \rho}=&\widehat{\mathcal{C}}^S_{\rho,3}RePr\left(\text{i}\Omega_2 \mathcal{I}-\widehat{{\nabla}}^2\right)^{-1}\widehat{\mathcal{B}}^S_{\rho,3}.
\end{align}
\end{subequations}
\endgroup
Taking the operation that $\|\cdot\|_{\infty}=\underset{\omega\in\mathbb{R}}{\text{sup}}\bar{\sigma}[\cdot]=\underset{\Omega_1\in\mathbb{R}}{\text{sup}}\bar{\sigma}[\cdot]=\underset{\Omega_2\in\mathbb{R}}{\text{sup}}\bar{\sigma}[\cdot]$, we obtain the scaling relation in theorem \ref{thm:stratified_scaling_Re_Pr}(a). 

Using the relation that $\mathcal{H}^S_{\nabla ij}=\widehat{\boldsymbol{\nabla}}\mathcal{H}^S_{ij}$ in equation \eqref{eq:stratified_H_nabla_ij} with $i=u,v,w,\rho$, and $j=x,y,z,\rho$, and similarly employ the notion of $\|\cdot\|_{\infty}$, we obtain the scaling relation in theorem \ref{thm:stratified_scaling_Re_Pr}(b). 
\end{myproof}

In figure \ref{fig:stratified_block_diagram_Re_Pr_scaling}, the block $-\text{i}k_zU'$ inside the dashed line ({\color{red}$\dashed$}, red) contributes to the relatively large scalings of $\|\mathcal{H}^S_{uy}\|_{\infty}\sim Re^2$, $\|\mathcal{H}^S_{uz}\|_{\infty}\sim Re^2$ at high $Re$ in equation \eqref{eq:stratified_scaling_Re_Pr_a} of theorem \ref{thm:stratified_scaling_Re_Pr}(a), which has been attributed to the lift-up mechanism; see discussion in \citet{jovanovic2020bypass}. Similarly, the block $-\overline{\rho}'$ outlined by ({\color{blue}$\dashdot$}, blue) contributes to the relatively large scalings of $\|\mathcal{H}^S_{\rho y}\|_{\infty}\sim Re^2Pr$, and  $\|\mathcal{H}^S_{\rho z}\|_{\infty}\sim Re^2Pr$ at high $Re$ or $Pr$. This similarity between streamwise streaks and density streaks is consistent with the observation that passive scalar streaks can be generated by the same lift-up mechanism as the streamwise streaks \citep{chernyshenko2005mechanism}.

\subsection{Proof of theorem \ref{lemma:stratified_mu_componentwise_inf}}
\begingroup
\allowdisplaybreaks
\begin{myproof}
We define the set of uncertainties:
\begin{subequations}
% \label{eq:stratified_uncertain_set}
\begin{align}
\mathbfsbilow{\widehat{U}}^S_{\Upxi,ux}:=\left\{\text{diag}\left(-\mathbfsbilow{\widehat{u}}_{\xi}^{\text{T}},\mathsfbi{0},\mathsfbi{0},\mathsfbi{0}\right):-\mathbfsbilow{\widehat{u}}^{\text{T}}_{\xi}\in \mathbb{C}^{N_y\times 3N_y}\right\},\label{eq:stratified_uncertain_set_u}\\
\mathbfsbilow{\widehat{U}}^S_{\Upxi,vy}:=\left\{\text{diag}\left(\mathsfbi{0},-\mathbfsbilow{\widehat{u}}_{\xi}^{\text{T}},\mathsfbi{0},\mathsfbi{0}\right):-\mathbfsbilow{\widehat{u}}^{\text{T}}_{\xi}\in \mathbb{C}^{N_y\times 3N_y}\right\},\label{eq:stratified_uncertain_set_v}\\
\mathbfsbilow{\widehat{U}}^S_{\Upxi,wz}:=\left\{\text{diag}\left(\mathsfbi{0},\mathsfbi{0},-\mathbfsbilow{\widehat{u}}_{\xi}^{\text{T}},\mathsfbi{0}\right):-\mathbfsbilow{\widehat{u}}^{\text{T}}_{\xi}\in \mathbb{C}^{N_y\times 3N_y}\right\},\label{eq:stratified_uncertain_set_w}\\
\mathbfsbilow{\widehat{U}}^S_{\Upxi,\rho\rho}:=\left\{\text{diag}\left(\mathsfbi{0},\mathsfbi{0},\mathsfbi{0},-\mathbfsbilow{\widehat{u}}_{\xi}^{\text{T}}\right):-\mathbfsbilow{\widehat{u}}^{\text{T}}_{\xi}\in \mathbb{C}^{N_y\times 3N_y}\right\},\label{eq:stratified_uncertain_set_rho}
\end{align}
\end{subequations}
Here, $\mathsfbi{0}\in \mathbb{C}^{N_y\times 3N_y}$ is a zero matrix with the size $N_y\times 3N_y$. 
Then, using the definition of the structured singular value in definition \ref{def:mu_stratified}, we have:
\begin{subequations}
\label{eq:stratified_mu_componentwise}
\begin{align}
    &\mu_{\mathbfsbilow{\widehat{U}}^S_{\Upxi,ux}}\left[\mathbfsbilow{H}^S_{\nabla}(k_x,k_z,\omega)\right]\nonumber\\
    =&\frac{1}{\text{min}\{\bar{\sigma}[\mathbfsbilow{\widehat{u}}^S_{\Upxi,ux}]\,:\,\mathbfsbilow{\widehat{u}}^S_{\Upxi,ux}\in \mathbfsbilow{\widehat{U}}^S_{\Upxi,ux},\,\text{det}[\mathsfbi{I}-\mathbfsbilow{H}^S_{\nabla}(k_x,k_z,\omega)\mathbfsbilow{\widehat{u}}^S_{\Upxi,ux}]=0\}}\label{eq:stratified_mu_componentwise_1}\\
    =&\frac{1}{\text{min}\{\bar{\sigma}[-\mathbfsbilow{\widehat{u}}^{\text{T}}_{\xi}]\,:\,-\mathbfsbilow{\widehat{u}}^{\text{T}}_{\xi}\in \mathbb{C}^{N_y\times 3N_y},\,\text{det}[\mathsfbi{I}_{3N_y}-\mathbfsbilow{H}^S_{\nabla ux}(k_x,k_z,\omega)(-\mathbfsbilow{\widehat{u}}^{\text{T}}_{\xi})]=0\}}\label{eq:stratified_mu_componentwise_2}\\
    =&\bar{\sigma}[\mathbfsbilow{H}^S_{\nabla ux}(k_x,k_z,\omega)]\label{eq:stratified_mu_componentwise_3}
\end{align}
\end{subequations}
Here, equality \eqref{eq:stratified_mu_componentwise_1} is obtained by substituting the uncertainty set in \eqref{eq:stratified_uncertain_set_u} into definition \ref{def:mu_stratified}. The equality \eqref{eq:stratified_mu_componentwise_2} is obtained by performing a block diagonal partition of terms inside of $\text{det}[\cdot]$ and employing zeros in the uncertainty set in equation \eqref{eq:stratified_uncertain_set_u}. Here, $\mathbfsbilow{H}^S_{\nabla ux}$ is the discretization of $\mathcal{H}^S_{\nabla ux}$ and $\mathsfbi{I}_{3N_y}\in \mathbb{C}^{3N_y\times 3N_y} $ in \eqref{eq:stratified_mu_componentwise_2} is an identity matrix with matching size $(3N_y\times 3N_y)$, where we use the subscripts to distinguish it from $\mathsfbi{I} \in \mathbb{C}^{12N_y\times 12N_y}$ in \eqref{eq:stratified_mu_componentwise_1}. The equality \eqref{eq:stratified_mu_componentwise_3} uses the definition of unstructured singular value; see e.g., \citep[equation (11.1)]{zhou1996robust}. 

Similarly, we have:
\begin{subequations}
\label{eq:stratified_mu_componentwise_vwrho}
\begin{align}
    \mu_{\mathbfsbilow{\widehat{U}}^S_{\Upxi,vy}}\left[\mathbfsbilow{H}^S_{\nabla}(k_x,k_z,\omega)\right]=\bar{\sigma}[\mathbfsbilow{H}^S_{\nabla vy}(k_x,k_z,\omega)],\\
    \mu_{\mathbfsbilow{\widehat{U}}^S_{\Upxi,wz}}\left[\mathbfsbilow{H}^S_{\nabla}(k_x,k_z,\omega)\right]=\bar{\sigma}[\mathbfsbilow{H}^S_{\nabla wz}(k_x,k_z,\omega)],\\
    \mu_{\mathbfsbilow{\widehat{U}}^S_{\Upxi,\rho\rho}}\left[\mathbfsbilow{H}^S_{\nabla}(k_x,k_z,\omega)\right]=\bar{\sigma}[\mathbfsbilow{H}^S_{\nabla \rho \rho}(k_x,k_z,\omega)].
\end{align}
\end{subequations}
Using the fact that $\mathbfsbilow{\widehat{U}}^S_{\Upxi}\supseteq \mathbfsbilow{\widehat{U}}^S_{\Upxi,ij}$ with $ij=ux,vy,wz,\rho\rho$ and equalities in \eqref{eq:stratified_mu_componentwise}-\eqref{eq:stratified_mu_componentwise_vwrho}, we have:
\begin{align}
     \mu_{\mathbfsbilow{\widehat{U}}^S_{\Upxi}}\left[\mathbfsbilow{H}^S_{\nabla}(k_x,k_z,\omega)\right]\geq\mu_{\mathbfsbilow{\widehat{U}}^S_{\Upxi,ij}}\left[\mathbfsbilow{H}^S_{\nabla}(k_x,k_z,\omega)\right]=\bar{\sigma}[\mathbfsbilow{H}^S_{\nabla ij}(k_x,k_z,\omega)].
     \label{eq:stratified_mu_componentwise_inequality}
\end{align}
Applying the supreme operation $\underset{\omega \in \mathbb{R}}{\text{sup}}[\cdot]$ on \eqref{eq:stratified_mu_componentwise_inequality} and using definitions of $\|\cdot\|_{\mu}$ and $\|\cdot \|_{\infty}$ we have:
\begin{subequations}
\begin{align}
    \|\mathcal{H}_{\nabla }^S\|_{\mu}\geq& \|\mathcal{H}_{\nabla ux}^S\|_{\infty},\;\; \|\mathcal{H}_{\nabla }^S\|_{\mu}\geq \|\mathcal{H}_{\nabla vy}^S\|_{\infty},\tag{\theequation a,b}\\
    \|\mathcal{H}_{\nabla }^S\|_{\mu}\geq& \|\mathcal{H}_{\nabla wz}^S\|_{\infty},\;\;
    \|\mathcal{H}_{\nabla }^S\|_{\mu}\geq \|\mathcal{H}_{\nabla \rho \rho}^S\|_{\infty}.\tag{\theequation c,d}
\end{align}
\end{subequations}
This directly results in inequality \eqref{eq:stratified_mu_larger_than_all_diagonal_component} of theorem \ref{lemma:stratified_mu_componentwise_inf}. 
\end{myproof}
\endgroup

\bibliography{main_stratified}
\bibliographystyle{jfm}

\end{document}